\centerline{\bf The Status and Programs of the New Relativity Theory }
\bigskip

\centerline{Carlos Castro}
\centerline{Center for Theoretical Studies of Physical Systems}
\centerline{Clark Atlanta University, Atlanta, GA. 30314}  

\bigskip
\centerline{ November, 2000 } 

\bigskip

\centerline { Dedicated to the memory of my parents } 

\bigskip

\centerline{\bf Abstract} 
\bigskip

A review of the most recent results of the New Relativity Theory is presented. 
These include  a straightforward derivation of the Black Hole Entropy-Area relation 
and its $logarithmic$ corrections; the derivation of the string uncertainty relations and 
generalizations ; ; the relation between the four dimensional gravitational conformal anomaly 
and the fine structure constant; the role of Noncommutative Geometry, Negative Probabilities 
and Cantorian-Fractal spacetime in the Young's  two-slit experiment. We then generalize the 
recent construction of the Quenched-Minisuperspace bosonic $p$-brane propagator in $D$ dimensions 
($AACS$ [18] ) to the full multidimensional case involving all $p$-branes :  
the construction of the Multidimensional-Particle propagator  in Clifford spaces ( $C$-spaces ) 
associated with a nested family of $p$-loop histories living in a target $D$-dim background spacetime . 
We show how the effective $C$-space geometry is related to $extrinsic$  curvature of ordinary spacetime. 
The  motion of rigid particles/branes is studied to explain the natural $emergence$ of classical spin. 
The relation among $C$-space geometry and ${\cal W}$, Finsler Geometry and ( Braided ) Quantum Groups 
is discussed. Some final remarks about the Riemannian long distance limit of $C$-space geometry are made.

\bigskip
\bigskip

\centerline{\bf 1. Introduction : The New Relativity Theory } 
\bigskip

Before we begin the present status and prsopects of the New Relativity Theory [1] 
we deem it very important to review in {\bf 1} the New Extended Scale Relativity and its postulates. 
This is all one needs to construct in {\bf 2}  
the  generalization of the Quenched-Minisuperspace bosonic $p$-brane propagator in $D$ dimensions [18] ; i.e 
the Multidimensional-Particle propagator in $C$-spaces ( Clifford ) associated with a nested family of $p$-loop histories 
living in a $D$-dimensional  target spacetime background. 
In {\bf 3} we discuss the derivation of the Black Hole entropy and its $logarithmic$  corrections directly 
from a $p$-loop Harmonic oscillator in $C$-space ( Clifford space ). 
In {\bf 4  } we review the derivation of the string uncertainty relations and corrections thereof due to $all$ $p$-brane values ranging from $ p = 0 $ to $p= \infty$. ; 
In {\bf 5} we show how the effective $C$-space geometry is related to $extrinsic$  curvature of ordinary spacetime. 
The  motion of rigid particle/branes is studied in relation to the natural $emergence$ of classical $spin$  from $C$-space.Some final remarks  are made about the relation that the $C$-space geometry bears to ${\cal W}$ Geometry, Finsler Geometry 
and ( Braided ) Quantum Groups.   
In {\bf 6} the calculation of $ 4 + \phi^3 $ ( $\phi$ is the Golden Mean $0.618...$ )  
as the average dimensions of the world is discussed and the relation between the four dimensional 
gravitational conformal anomaly, Fractal Spacetime  and the fine structure constant is explicitly shown. 
The role of Noncommutative Geometry, Negative Probabilities 
and Cantorian-Fractal spacetime in the Young's  two-slit experiment of an indivisible quantum particle.     
is derived in {\bf 7} . 
In {\bf 8} we briefly discuss Quantum Groups, and the Braided Hopf Quantum Clifford Algebra associated with the 
Master Action Functional for the quantum dynamics of the Master Field in $C$-space.         
Finally, we present our conclusions. Some final remarks about the Riemannian long distance limit of $C$-space 
geometry are made.

\bigskip

\centerline{\bf 1.1 Historical Background} 
\bigskip

Recently we have proposed that a New Relativity principle may be 
operating in Nature
which could reveal important clues to find the origins of $M$ theory  
[1]. 
We were forced to introduce this New Relativity principle, where all 
dimensions and
signatures of spacetime are on the same footing, to find a fully 
covariant formulation
of the $p$-brane Quantum Mechanical Loop Wave equations. This New 
Relativity Principle,
or the principle of Polydimensional Covariance as has been called by 
Pezzaglia,
has also been crucial in the derivation of Papapetrou's equations of 
motion of a
spinning particle in curved spaces that was a long standing problem
which lasted almost 50 years  [2]. A Clifford calculus was used where 
all the
equations were written in terms of Clifford-valued multivector 
quantities;
i.e one had to abandon the use of vectors and tensors and replace them 
by
Clifford-algebra valued quantities, matrices, for example .

The New Extended Scale Relativity Theory has already allow us to derive, from first principles,  
the String Uncertainty 
Relations,
corrections thereof, and the precise connection between the Regge trajectory behaviour of the string spectrum and the {\bf area} quantization . 
 The full blown infinite-dimensional Quantum Spacetime Generalized Uncertainty relations that included the contributions 
of {\bf all} $p$-branes, not only due to strings,  was given [3]. In [4] we were able to show that there are {\bf no} such things as {\bf EPR} paradoxes in this 
New Scale Relativity. The latter is truly a Machian Relativity theory, were relationships are the only meaningful statements one can make. 
Originally it was formulated in an abstract Categorical space, {\bf C}-spaces [1]. 
The New Relativity sprang out from Laurent Nottale's Scale Relativity [5]; 
Mohammed El Naschie Transfinite Cantorian-Fractal Spacetime [6]; Garnet Ord's original work on 
fractal random walks [7] ; the classic Geoffrey Chew's Bootstrap hypothesis : all {\bf p}-branes are made of each other  
( William Pezzaglia's principle of polydimensional covariance ); and 
$p$-adic Physics and  Non-Archimedean Geometry by Siddarth, Pitkannen, Khrennikov, Freund, Volovich, Vladimorov, Zelenov and many others [8].

The goals and partial list of sucesses of the New Relativity, formulated in Cantorian-Fractal Spacetime, 
are   : 

{\bf 1} - Hope to provide with a truly background independent formulation of Quantum Gravity and $M$ theory. 

{\bf 2}- To furnish 
the physical foundations  underlying String and $M$ theory. 
So far nobody has been able to answer the question : " What {\bf is} String Theory ? ". The New Relativity may 
provide with a plausible answer. In particular we were able to derive the string uncertainty relations directly from 
the wave equations of the New Relativity [10] and its $p$-brane generalizations. 

{\bf 3}- Prove why we live in $3+1$ dimensions. In [9] we have shown that the average dimension of the world is of the order of $4 + \phi^3 $. 

{\bf 4}- Solve the cosmological constant problem [9] , 
in a similar fashion that Nottale did in his initial work on Scale Relativity. And discuss of the existence of a 
dimensional phase transition from $4 + \phi^3 $ to $\phi^3 $ , the so called Noncommutative quasi-crystal phase. 

{\bf 5}- Provide with a physical meaning to negative topological dimensions and to the notion of negative-information entropy, or anti-entropy as has been called by M. Conrad. . 

{\bf 6}- Write down the Unique Quantum Master Action functional of the world, in an abstract {\bf C}-space, outside spacetime, a generalized Twistor space, that governs the quantum dynamics 
for the creation of spacetime itself; gravity and all of the other fundamental forces in Nature.

{\bf 7}- Propose  a solution to how chiral symmetry-breaking occurs in Nature and much, much more [10]. 
\smallskip

With this preamble let us summarize briefly what are the basic postulates of the New Relativity Theory next. 
This will allow us to explain in full rigour why the fundamental constants in Nature, like Planck's constant, 
change with " time ";  i.e. with the Renormalization Group flow of the microscopic ( scaling ) arrow of time: with energy or resolutions. 
We will explain why there is an effective value of the Planck constant when one 
approaches scales comparable with the Planck length : when quantum gravitational phenomena are important.  

\bigskip
\centerline{\bf 1.2 The Postulates of the New Scale Relativity Theory} 

\bigskip

{\bf 1} . Physics is an experimental science. Physics is about measurements. In order to perform a measurement one needs a standard of measure to compare 
measurements with. In Einstein's Relativity he introduced the speed of light as the velocity standard to compare velocities with. 
From the mathematical  point of view one can say : let us have a field at {\bf x}. 
To define the real number {\bf x}, mathematically, requires knowing its value with infinite precision : 
simply adding digits at will. To do it physically is another story. 
It would require a computer an infinite amount of memory just to store all this infinite amount of information. 
One way to avoid the problem of using real numbers is to 
introduce $p$-Adic numbes in Physics [8]. 

For these reasons, Laurent Nottale [5] introduced his original Scale Relativity by postulating that the Planck scale is the universal standard of measurements. It is 
the minimum, impassible distance in Nature, 
invariant under scale-relativistic dilatations. Exactly along similar lines, Einstein's motion Relativity was based on taking the speed of light as the maximum, and {\bf invariant},  
attainable speed in Nature.  
The Planck scale in four dimensions is given in terms of the $3$ fundamental constants, speed of light $c$ , 
Newton's constant in four dimensions $G$ and Planck's  constant $\hbar $ :

$$\Lambda = \sqrt { {\hbar G \over c^3 }} = 10^{-33} cms . \eqno (1)$$

Notice that if $\hbar =0$, $G =0$,  $c =\infty$ the Planck scale would have been automatically zero. Meaning that one would not have  "quantum mechanics " ; the gravitational force will reduce to $0$ and there would not be " Lorentz invariance ", only Galilean symmetry. The invariant minimum scale in all dimensions, $\Lambda$ will be set to 
{\bf unity}; in units where $\hbar = c = 1$. The Planck scale is explictly dimension dependent through its dependence on the Newton constants.  
So is the Newton gravitational  constant. For example, in $D, D-1, D-2,....$ , the Planck scale ( that we set to unity) is given in terms of the Newton constants :

$$\Lambda = G_D^{(1/D-2)} = G_{D-1}^{(1/D-3)} = ......= 1 .\eqno (2a)$$
Taking  logarithms ( in any base if one wishes )  on eq-(2a),  one has the relationship among the Newton constants, in $D$ and $D-1$ diemnsions, 
which does {\bf not} require any compactifications whatsoever, as it is assumed in conventioanl string and Kaluza-Klein supergravities, : 
$${D-3 \over D-2} = { ln~G_{D-1} \over ln~G_D} = { log~G_{D-1} \over log ~G_D }.                     \eqno (2b)$$
Notice that when $D = 2$, the Newton constant in two-dimensions is set to $G_2 = 1$ so that $1^\infty =1$. 
For example in $D =2, G_2 =1 \rightarrow ln~G_2 = ln~1 = 0$ which is consistent with all the denominators of eq-(2b). 

Variable speed of light Cosmologies are becoming very popular today. This does not imply that Einstein was wrong at his time. 
This only means, as Dirac pointed out long ago, that the 
fundamental constants can change slowly with " time " : the value of the constants slowly flow with the Renormalization group, from the Ultraviolet ( small scales ) to the Infrared ( large scales ). The speed of light when Einstein formulated his theory is the same as today. Only during the early Universe there were substantial changes.  
The constants flow with the provision that the truly fundamental " relativistic " invariant in Nature, the Planck scale, remains fixed. 
Eq-(1) by inspection, entails that 
$\hbar, c, G$ could flow with the RG flow in such a fashion that they will leave $\Lambda$ invariant.  
This explains how, at very large energies ( Planckian ) , at very small scales ( Planck scales), we can begin to see the corrections to the fundamental constants. We have now 
an understanding as to why one has an effective Planck constant that can vary with energy once we approach scales comparable to the Planck scale. 
Exactly the same thing happens when we approach the speed of light : the masses begin to grow compared to their rest mass values.  

\smallskip

{\bf 2} . The Principle of Poly-dimensional covariance and the Clifford-algebra-multivector calculus. 

The New Relativity is a true Machian one. Relationships are the  only meaningful statements one can make. 
To view a single $p$-brane as an isolated entity is a meaningless concept. $p$-branes are solely defined in terms of others. For this reason, 
we included the Chew boostrap hypothesis as a crucial ingredient, and wrote : 

{\bf All} $p$-branes are made of each-other. This is, in essence,  the origins of the dualities in $M$ theory. 
Pezzaglia [2] using Chew's bootstrap hypothesis coined the term 
" poly-dimensional-covariance ". Since now we have all $p$-branes, of all dimensionalities from $ p =0, 1, 2....\infty$, 
the poly-dimensional covariance is the statement that 
all $p$-branes ( dimensions ) rotate into each-other. Exactly what happens with ordinary Lorentz transformations : 
the axis are entangled and space and time are mixed. For the role of objects of negative topological dimensions see [10]. 
For example, in ordinary string theory, an object of dimension $p =- 1$ spans a $p+1 =  -1 + 1 = 0$-dimensional  "worldline"  : it is an instanton.

Space and time have different units. Einstein was able to connect them by the introduction of an invariant  velocity parameter : 
the speed of light. 
If one sets $c = 1$ in Einstein's  theory, space and time have the same " units " : they are exchangeable. 
Since the Planck scale has units of length, upon setting $\Lambda = 1$, it means that all dimensions are 
are exchangeable also. Dimensions then can be " rotated " into each-other. $p$-branes can be transformed into each-other ( duality principle ).  
Einstein's Relativity required the use of a Lorentz four-vector to embrace space with time; energy with momentum, etc....
In the New Relativity one needs Clifford-algebra-valued multivectors. The latter multi-vector is a mathematical object that encodes 
all objects of different dimensionalities. It encodes all $p$-brane histories,  embedded in a target spacetime background,  into one single scoop. 
For the analog of ordinary Lorentz transformations that rotate space into time, for example, one has polydimensional transformations, that rotate $p$-branes among each-other. 

For this reason we will write our New Relativistic wave equations in a Clifford-space, 
whose derivative elements will be {\bf quadratic}  and have, in addition to the  ordinary quadratic derivatives with respect to ordinary 
point coordinates, $x^\mu$, quadratic derivatives w.r.t the holographic area-coordinates, holographic volume-coordinates, ..... 
We will explain this in detail in section {\bf 2} what these holographic area, volume, hypervolume....coordinates really are.

{\bf 3} . Einstein's General Relativity required a Riemannian Geometry; 
the New Relativity requires a Cantorian-Fractal Spacetime model developed by Mohammed El Naschie [6]. 
The latter is an example of Von Neuman's PointlessNoncommutative Geometry. The world is 
multi-fractal. For a detailed analyis of Noncommutative geometry, etc...see [10] and references therein.  

Essentially one has that because the Planck scale is the minimum distance in Nature; there are no such things as " points " in Nature. Only in Mathematics. 
 A " point " is smeared into a fuzzy ball of all 
possible topological dimensision. On average,  the " points " are four-dimensional spheres, for this reason we live effectively in four-dimensions : we perceive 
an average dimension 
taken over all the infinite-dimensional spacetime [10]. This occurs exactly in the same fashion that we only measure the average velocity of an 
ensemble of molecules in a room, 
when we measure the temperature of the room, pressure, etc...Cantorian-Fractal Geometry is a " pointless " Noncommutative Geometry. As we zoom into what looks like a 
" point" from a distance, we realize that it is a four-dimensional sphere. As we zoom in deeper into what we thought looked like a point inside 
that small sphere we realize that it is a smaller four-dimensional sphere; and so forth ad infinitum. We cannot reach the Planck scale. 
Like a massive object cannot reach the speed of light.  Nature is multifractal.

{\bf 4} Non-Archimedean geometry and $p$-Adic Physics [8]

Since the Planck scale is the minimum attainable scale in Nature, Nottale provided with the scaling analogs of Lorentz transformations : The composition of two 
scalings ( contractions ) cannot yield scales smaller than the Planck scale. In the same way that the addition of two velocities cannot exceed the speed of light.
This forces one to abandon the Archimedean Geometry for a Non-Archimedean one and the replacement of real numbers for $p$-Adic ones [8]. Using a $p$-Adic norm, 
which allows to define $p$-adic numbers based on another extension of the rational numbers , one can show that the net resulting $p$-adic norm, of the composition/sum of two 
$p$-adic numbers,  is less than the $p$-adic norm of the larger norm of the initial two $p$-adic numbers. 
Roughly speaking, the $p$-adic norm obeys an ultra-metricity condition. $p$-Adics are the natural numbers to use in the New Relativity. The latter Relativity is consitent with a Non-Archimedean Geometry.     
\bigskip

What we find most important in using primes, the atoms of numbers, in the New Relativity is that 
all prime numbers must appear on equal footing. In the same way that all dimensions did. 
Since dimensions change this automatically entails that
one must include also all topologies on equal footing : a `` Topological Relativity'' 
as has been called by Finkelstein and others.
$p$-Adic numbers have been used to label topologies. Not surprisingly, this fits very well within this new framework.   
It is also well known to the experts that there is a very deep connection between Quantum groups and 
$p$-Adic numbers when the deformation parameter $ q = 1/ p$. The Real number limit $ p = \infty$ 
is equivalent to the classical group limit 
$ q = 0$. Quantum Symmetric spaces  `` interpolate `` between the Real numbers and the $p$-Adic ones.

To sum up, the world is {\bf not } scale invariant. Dimensions are in the eye of the beholder [1,2]. They are resolution-dependent [5,6].  
As we probe into smaller regions, larger energies, with the "microscope" of the RG flow, more dimensions are accesible to us. 
Things look different to two observers living 
in two different scale-frames of references. The most famous example : it is meaningless to compare the vacuum energies in 
two completely different scale-frames of references : the Planck scale and the Hubble scale. This is the fundamental reason why the " cosmological" 
constant differs by $60$ orders of magnitude : 
the so-called cosmological constant " problem ". This inconsistency, was elegantly solved by Nottale [5] and recently by us [9] within the context of 
Renormalization Group techniques and self-organized 
non-equlibrium critical phenomena in Cosmology, intially emphasized by Smolin and Kauffmann [19].

Based on this cursory introduction to the basic principles of the New Relativity, we can understand now why the Planck constant $\hbar$, for example, is not a true constant, 
but it varies with energy, resolution. Once, of course, we approach the Planck scale. At ordinary energies, $\hbar$ is a constant. 
This is why these effects have not been detected in
ordinary experiments. The energy is extremely low in comparison with the Planck energy of $10^{19}$ GeV. 

\bigskip
.

\centerline{ \bf 2. Quenched-Minisuperspace Bosonic $p$-brane Propagator and its C-space generalization} 
\bigskip

Recently $AACS$ [18] we were able to write down the Quenched-Minisuperspace Bosonic $p$-brane 
propagator by borrowing the Minisuperspace approximation from Cosmology, and the `` quenching `` procedure from $QCD$. This new approximation provided an 
$exact$ description of both the $collective~mode $ deformation of the brane and the center of mass dynamics in the target spacetime. Earlier work based on $loop$ spaces allowed [30] to write down the Schwinger-Fock proper time formulation of 
  wiggled  $p$-branes.

Imagine one has a family of $p$-branes moving in a target spacetime background, where the values of $p$ range from $p=0$, a point history; $p =1$ a closed string history; $p =2$ a closed membrane history. A membrane of topology of a sphere, for example : a {\bf 2}-loop; and so forth. 
Until one saturates the spacetime with the spacetime filling $p$-brane : $ p+1 = D$.  
The family of $p$-brane degrees of freedom are encoded in term of hyper-matrix coordinates [1]. .

Generalized " hyper-matrix coordinates " transformations in the New Relativity reshuffle, for example, 
a loop history into a membrane history; a membrane history into a
into a $5$-brane history; a $5$-brane history into a $9$-brane history  and so forth; in particular it can transform a $p$-brane history into suitable combinations 
of other $p$-brane histories as building blocks. This is the bootstrap idea taken from the point particle case to to the $p$-branes case : 
each brane is made out of all the others. " Lorentz" transformations in {\bf 
C}-spaces
involve hypermatrix changes of " coordinates " [1] . The naive Lorentz 
transformations do not
apply in the world of Planck scale physics. Only at large scales the 
Riemannian continuum is
recaptured . For a discussion of the more fundamental Finsler Geometries implementing the minimum scale ( maximal proper acceleration ) in String 
Theory see [13].

There was a one-to-one correspondence between the nested hierarchy of 
point, loop, {\bf 2}-loop,
{\bf 3}-loop,......{\bf p}-loop histories encoded in terms of 
hypermatrices [1] 
and wave equations written in terms of Clifford-algebra valued 
multivector quantities.[2] 
This permitted  us to recast the QM wave equations associated with the 
hierarchy of nested 
{\bf p}-loop histories, embedded in a target spacetime of $D$ dimensions 
, 
where the values of $p$ range  from :  $p=0,1,2,3......D-1$, as a 
$single$ QM {\bf line} 
functional wave equation whose lines live in a Noncommutative Clifford 
manifold of
$2^D$ dimensions. $p=D-1$ is the the maximum value of $p$ that 
saturates the
embedding spacetime dimension. 

An action {\bf line}  functional, associated with the {\bf interacting} QFT of {\bf lines} in 
Noncommutative Clifford manifolds {\bf C}-spaces,  was launched forward in [1]. The QFT program of such interacting field theory of {\bf C}-lines in Noncommutative spaces, 
a generalized Twistor theory, is currently under investigation [18]. One will have cubic interactions  
associated with the product and coproduct of a Braided-Hopf-Quantum-Clifford algebra [15]. 
The product represents the annihilation of two {\bf C}-lines into a third one. The coproduct represents the creation of two lines from one line . The 
quartic interactions correspond to the {\bf braiding} of two-lines into another two-lines : scattering. One has here more complicated statistics than the ordinary 
bose/fermions one : it is a braided one !. The two-point vertex corresponds to a pairing of the algebra 
representing the composition of two lines  into a $0$-line. The kinetic terms are the extensions of 
Witten-Zwiebach open/closed string field theory, based on the Batalin-Vilkovisky Quantum Master action [16,17] . 
The closed-string field theory action required the used of Operads and Gerstenhaber algebras [17].   
Such QFT in Noncommutative spaces is a very complex one due to the Ultra-Violet/Infrared entanglement. 
Due to string duality, there is a maximum scale dual to the minimum Planck scale.
In [9] we provided with integral expressions to determine the maximum scale  that is dual to the Planck scale.

The line functional wave equation in the Clifford manifold, {\bf 
C}-space, for the simplest `` linear `` case   is :

$$\int d\Sigma~( {\delta^2 \over \delta X(\Sigma) \delta X(\Sigma) } 
+{\cal E}^2 )
\Psi [X(\Sigma)]=0. \eqno (3)$$
where $\Sigma$ is an invariant evolution parameter of $l^{D}$ dimensions 
generalizing the notion of the invariant proper time in Special 
Relativity
linked to a massive point particle line ( path ) history :

$$(d\Sigma)^2 = (d\Omega_{p+1})^2 + \Lambda^{2p}(dx^\mu dx_\mu)
+  \Lambda^{2(p-1)}(d\sigma ^{\mu\nu}  d\sigma_{\mu\nu} )
+  \Lambda^{2(p-2)}(d\sigma^{\mu\nu\rho} d \sigma _{\mu\nu\rho})
+ .......\eqno (4)$$
$\Lambda$ is the Planck scale in $D$ dimensions : $\Lambda = G_D^{{1\over D-2}}$ where $G_D$ is Newton's constant in $D$ dimensions.  
{\bf X}$(\Sigma)$  is a Clifford-algebra valued " line " living in the 
Clifford
manifold ( {\bf C}-space)  :

$$X=\Omega_{p+1} +\Lambda^p x_\mu \gamma^\mu +\Lambda^{p-1} \sigma_{\mu\nu} \gamma^\mu 
\gamma^\nu +\Lambda^{p-2}
\sigma_{\mu\nu\rho}\gamma^\mu \gamma^\nu \gamma^\rho +......... \eqno 
(5a)$$

The multivector {\bf X} encodes in one single stroke the point history 
represented by the
ordinary $x_\mu$ coordinates and the holographic projections of the 
nested family of   
{\bf 1}-loop, {\bf 2}-loop, {\bf 3}-loop...{\bf p}-loop histories onto 
the embedding
coordinate spacetime planes given respectively by : 
$$\sigma_{\mu \nu},  \sigma_{\mu \nu\rho}......\sigma_{\mu_1 
\mu_2...\mu_{p+1}}\eqno (5b)$$
The scalar $\Omega_{p+1}$ is the invariant proper 
$p+1=D$-volume associated
with the motion of the ( maximal dimension ) {\bf p}-loop across the 
$D=p+1$-dim target
spacetime. It naturally couples to the unit matrix of the Clifford algebra.

There was a coincidence condition [1] that required to equate the values 
of the
center of mass coordinates $x_\mu$, for all the {\bf p }-loops, with the 
values of the
$x^\mu$ coordinates of the
point particle path history. This was due to the fact that
upon setting $\Lambda=0$ all the {\bf p}-loop histories collapse to a 
point history. 
The latter history is the baseline where one constructs the whole 
hierarchy. 
This also required a proportionality relationship :

$$\tau \sim { A\over \Lambda }\sim {V \over \Lambda^2}\sim.......
\sim {\Omega_{p+1} \over \Lambda^p}. \eqno (6)$$
$\tau,A,V....\Omega^{p+1}$ represent the invariant proper time, proper 
area, proper volume,...
proper $p+1$-dim volume swept  by the  point, loop, {\bf 2}-loop, 
{\bf 3}-loop,.....
{\bf p}-loop histories across their motion through the embedding spacetime, respectively.
${\cal E}=T $ is a quantity of dimension $(mass)^{p+1}$, the maximal 
$p$-brane tension ( $p=D-1$) .

A {\bf C}-line in {\bf C}-space is nothing but a Clifford algebraic extension of Penrose's twistors. From a distance, the line looks like a point 
history : a one-dimensional world line. Upon closer inspection, upon zooming in, we realize that it corresponds to the center-of-mass motion of a closed string history, a 
{\bf 1}-loop. And that the line turns into a two-dimensional surface : the lateral area-swept by the closed-string. 
Upon a further inspection, as we zoom in deeper, we realize that the closed-string history is really a closed-membrane history; and so forth and so forth.
Dimensions are resolution and energy dependent. All these $p$-brane histories that have a common center-of-mass coordinate , 
are encoded in terms of the Clifford-algebra-valued 
{\bf C}-lines , a generalized twistor. The holographic, or shadow-projections,  of the areas, volumes, hypervolumes,....onto the respective 
coordinate planes are nothing but the holographic coordinates of the {\bf C}-lines.

The wave functional $\Psi$ is in general a Clifford-valued, hypercomplex 
number.
In particular it could be a complex, quaternionic or octonionic valued 
quantity.
At the moment we shall not dwell on the very subtle complications and 
battles associated
with the quaternionic/octonionic extensions of Quantum Mechanics [14] 
based on Division algebras and simply take the wave function to be a 
complex number. 
The line functional wave equation for lines living in the Clifford 
manifold ( {\bf C}-spaces)
are difficult to solve in general. To obtain the Bekenstein-Hawking Black-Hole Entropy-Area 
relations, 
and corrections thereof,  one needs to simplify them.

The most simple expression ( all modes are frozen except the zero modes ) is to write the simplified wave equation 
for the family of free ( non-interacting ) $p$-loops in $D$-dimensions  , in units 
$\hbar=c=1$ :

$$\{ ~ - {1\over 2 \Lambda^{ p -1} }  [  {\partial ^2 \over \partial  x^\mu  \partial  x_\mu }
+ {\Lambda^2 }  {\partial ^2 \over \partial \sigma^{\mu\nu} 
\partial  \sigma_{\mu\nu}} + {\Lambda^4  } {\partial ^2 \over \partial \sigma^{\mu\nu\rho} 
\partial \sigma_{\mu\nu\rho} }   +......] \} ~ \Psi =     
T ~ \Psi [x^\mu, \sigma^{\mu\nu}, \sigma^{\mu\nu\rho},.....  ]. \eqno 
(7)$$
where $T$ is the tension associated with the maximal spacetime filling $p$-brane : $ p+1 = D$. It has units of energy per unit $p$-Volume ; i.e $(mass)^{p+1}$. 

Following the result for the ordinary point-particle propagator $K(x_b, x_a; \tau_b - \tau_a  )$ obeying :

$$ - { 1\over 2m} {\partial^2 \over \partial x^2_a } K(x_b, x_a; \tau_b - \tau_a  ) = i {\partial \over \partial \tau_a}
K(x_b, x_a; \tau_b - \tau_a  ). \eqno (8)  $$
whose solution is :

$$K(x_b, x_a; \tau_b - \tau_a  )= ( {m \over 2\pi (\tau_b - \tau_a)})^{{1\over 2}}
~exp~[ { 1/2 ~i m (x_b - x_a )^2 \over \tau_b - \tau_a}]. \eqno (9)$$

One can notice the role of the quantities $m$ ( mass )  and $\tau$ ( proper time ) in the expression for the kernel. 
One can generalize this result to $C$-spaces by finding the analog of a $C$-space invariant; i.e a 
polydimensional invariant parameter of dimensions of length $\lambda$ cosntructed out of the two $C$-space 
invariants : The $C$-space analog of proper time interval $\Sigma$ given in eq-(4) and the analog of `` mass `` : the quantity $m_{p+1}$ obeying the on shell condition for the 
polydimensional Clifford-algebra valued momentum : 

$$P^2 = ({1\over \Lambda})^{2p} (p_\mu p^\mu ) +  
({1\over \Lambda})^{2p-2}(p_{\mu\nu} p^{\mu\nu}) +   ({1\over \Lambda})^{2p-4}(p_{\mu\nu\rho} p^{\mu\nu\rho})              ........... + ( \mu_o )^2        =        ( m_{p+1}      )^2. \eqno (10) $$
where the canonical conjugate variable to the worldvolume $\Omega_{p+1}$ 
of the maximal spacetime filling $p$-brane ( $p+1 = D$) is nothing but the $cosmological ~constant ~\mu_o $  
of dimensions $ ( mass )^{p+1}$.  
Because the cosmological `` constant `` $\mu_o$ is itself a component of the Clifford-algebra valued polymomentum $ P$ this means that the cosmological `` constant `` $is ~not $ a $C$-space invariant by definition ! Its value can $rotate $ under polydimensional rotations ! [3, 9 ]

The natural $C$-space invariant quantity ${\cal L}$ of dimensions of $(length)^{p +1}$ allows to define a natural length scale 
$\lambda$ , which is just nothing but the analog of the $Fock-Schwinger-Feynman$ formulation based on the 
auxiliary parameter $\lambda$ [18, 30]  : 

$${\cal L}^2 \equiv { \Sigma_{p+1}\over m_{p+1} } \Rightarrow \lambda = {\cal L}^{(1/p+1)} = 
({ \Sigma^{p+1}\over m_{p+1} })^{ (1/ 2p+2) }. \eqno (11)$$ 
that involves the analog of mass : $m_{p+1}$ and the analog of proper time $\Sigma_{p+1}$ for the 
Multidimensional-Particle. These quantities are genuine invariants under automorphisms of the Clifford algebra ; i.e 
polydimensional rotations or automorphisms of the Clifford algebra basis `` vectors `` $\gamma^\mu $ [2] in such a 
way that the Clifford valued multivector $X$ in eq-(5a) rotates as follows : 

$$ \gamma^{\mu} \rightarrow \Gamma^{-1} ~ \gamma^\mu ~\Gamma. ~~~ X \rightarrow M X \eqno (12) $$
where $M$ is a $ 2^D \times 2^D $ matrix. One must arranged the $2^D$ components of the Clifford algebra 
valued 
multivector $ X$ : $\Omega_{p+1}; x_\mu, \sigma_{\mu\nu},  \sigma_{\mu\nu\rho}         ...$ 
into a column matrix of $2^D$ entries. The polydimensional rotations will $reshuffle $ the components/dimensions in 
such a way that a membrane history rotates into a five-brane history; a nine-brane rotates into an eleven-brane history and so forth, or even mixtures of all brane histories . This is the $p$-brane  Bootstrapping generalization 
of Chew's particle bootstrap ideas extended to 
dimensions : Pezzaglia's polydimensional covariance principle.

Hence, these polydimesnional rotations are the extensions of ordinary Lorentz transformations 
leaving the $C$-space interval invariant : 

$$ ( d {\tilde \Sigma}  )^2 = (d {\tilde \Omega}_{p+1} )^2 + \Lambda^{2p} (d {\tilde x}^\mu d {\tilde x}_\mu ) 
+  \Lambda^{2p -2 } (d {\tilde \sigma}^{\mu\nu}  d {\tilde \sigma}_{\mu\nu }  ) + ..................                          = $$
$$( d\Sigma )^2 = (d{\Omega}_{p+1} )^2 + \Lambda^{2p} (d { x}^\mu d {x} _\mu ) 
+  \Lambda^{2p -2 } (d { \sigma}^{\mu\nu}  d { \sigma}_{\mu\nu }  ) + ................... \eqno (13) $$
                   
The double covering of ordinary Lorentz transformations, in spinorial notation,  
can be seen as $ SL ( 2, C) $ rotations/Mobius transformations using $2 \times 2$ complex valued 
matrix whose entries  are respectively $(a, b, c, d )$ : 

$$ z \rightarrow {az + b \over cz + d }.~~~ ad - bc = 1. ~~~$$
This is attained by assigning to each four vector $x^\mu = ( x^0, x^1, x^2, x^3 ) $ 
the $2 \times 2 $ matrix $X$ using  
the three Pauli spin matrices $\sigma^i$ and unit matrix $I$ : 

$$ x^\mu = ( x^0, x^1, x^2, x^3 ) \leftrightarrow X= 
x^0 I + x^1\sigma_1 +  x^2\sigma_2 +  x^3\sigma_3 \Rightarrow  A^{-1} X A = 
{\tilde X}  = {\tilde x}^0  I + {\tilde x}^1\sigma_1 +  {\tilde x}^2\sigma_2 +  {\tilde x}^3\sigma_3. \eqno (14) $$
where $ A$ is an $SL(2,C)$ matrix. In this fashion one obtains the Lorentz transformations of 
${x}^\mu $ preserving the Lorentz norm $x^2_0 - x^2_1 - x^2_2 - x^2_3 $ of the four-vector $x^\mu$. ; 
i.e it is just  given by taking the trace 
$ Tr ( X^2 ) = Tr ( {\tilde X}^2 )$ due to its cyclic property.   

Based on the one-to-one correspondence bewteen $hypermatrices $ and Clifford algebra valued multivectors $X$ ( 5a ) 
one can extend the Lorentz transformations, using Pauli spin matrices , as polydimensional rotations or 
automorphisms of the Clifford algebra , that can be recast as automorphisms of $hypermatrices$, similarily to what 
occurs with ordinary Lorentz transformations expressed in terms of $SL(2, C) $ transformations of the 
$2 \times 2$ matrix $X$ given by eq-(14) . The analog of Null lines ( photons ) in $C$-space naturally correspond to 
$tensionless$ $p$-branes : $ m_{p+1} = 0$. Instead of null lines one has null tubes.

For example, in $ D = 4$ the degree of the Clifford algebra is $ 2^4 = 16 $ , meaning that one has $ 16$ independent 
$ 4 \times 4$ matrices : $ I, \gamma^\mu, \gamma^\mu \wedge \gamma^\nu,...$ spanning the Clifford algebra basis. 
These $16~4\times 4 $ matrices can be arranged as $4$ $cubic ~hypermatrices$  $ 4 \times 4  \times 4 $ : 
$ Y^0, Y^1, Y^2, Y^3$ where a similar type of transformation as eqs-(12, 14 ) 
via automorphisms of the Clifford algebra will map these $4$ $cubic$ hypermatrices 
into a new set ${\tilde Y}^0, {\tilde Y}^1,  {\tilde Y}^2, {\tilde Y}^3$. 
The generalization of Mobius transformations for $R^n= R^{ 2^D} $ 
can be obtained via the Vahlen matrices and the Vahlen group [25] . 
Whatever prescription one wishes to take, the idea is essentially to rotate the Clifford algebra multivector $X$ , with 
$2^D$ independent components,  by right multiplication with a $ 2^D \times 2^D $ matrix : 
$ {\tilde X } = M X $ leaving invariant the $C$-space interval or $norm $ of the $X$ Clifford algebra valued multivector.  
The matrix $M$ should obey the analog of the orthogonal/unitary matrix  
$M^T = M^{ -1} $ and $ M^+ = M^{ -1 }$ respectively.

Having discussed polydimensional rotations, autmorphisms of the Clifford algebra that leave the $C$-space inteval 
invariant, the invariant parameter $\lambda$ will allow us to extend the point particle kernel  to the $C$-space case by counting the number of degrees of fredom associated with the $collective$ excitations : there are $D$ degrees of freedom associated with the 
center of mass motion. There are $D(D-1)/2$ degrees of freedom associated with the $holographic$ area 
excitations $\sigma_{\mu\nu} $. There are  $D(D-1)(D-2)/6 $ degrees of freedom associated with the $holographic$ area 
excitations $\sigma_{\mu\nu\rho} $.           
 and so forth. The kernel in $C$-space can then be recast in terms of the analog of the 
Fock-Schwinger-Feynman parameter 
$\lambda$ [18, 30] and factorizes as follows :

$$ {\cal K} ~( X_a, X_b; ~\Sigma_b - \Sigma_a ) = K_{x^\mu} ( x^\mu_a, x^\nu_b; \Sigma_b - \Sigma_a  )  
~K_{\sigma^{\mu\nu} }( \sigma^{\mu\nu}_a,  \sigma^{\mu\nu}_b; \Sigma_b - \Sigma_a  )~ 
 K_{\sigma^{\mu\nu\rho} }( \sigma^{\mu\nu\rho}_a,  \sigma^{\mu\nu\rho}_b; \Sigma_b - \Sigma_a  )~                                              ...... \eqno (15)$$
where

$$K_{x^\mu} = ( {1 \over \lambda^2 })^{{D\over 2}}
~exp~[ { 1/2 ~i (x^\mu _b - x^\mu_a )^2 \over \lambda^2 } ]. \eqno (16)$$

$$K_{\sigma^{\mu\nu} }= ( {1 \over \lambda^4 } )^{{ D(D-1)\over 2.2  } }
~exp~[ { 1/2 ~i ( \sigma^{\mu\nu}_b - \sigma^{\mu\nu}_a )^2 \over \lambda^4 } ]. \eqno (17)$$

$$K_{\sigma^{\mu\nu\rho} }= ( {1 \over \lambda^6 } )^{{ D(D-1)(D-2)\over 2.2.3  } }
~exp~[ { 1/2 ~i ( \sigma^{\mu\nu\rho}_b - \sigma^{\mu\nu\rho}_a )^2 \over \lambda^6 } ]. \eqno (18)$$

...............

The $C$-space Kernel/Propagator  $ {\cal K} ~( X_a, X_b; ~\Sigma_b - \Sigma_a )$ is what allows to evaluate the wavefunction at two separate locations $X_a, X_b$ in a time span of $\Sigma_b - \Sigma_a$ :

$$\Psi ( x^\mu_b, \sigma^{\mu\nu}_b, \sigma^{\mu\nu\rho}_b..... ) = \int dx^\mu_a~d\sigma^{\mu\nu}_a~d\sigma^{\mu\nu\rho}_a.....d\Omega_{p+1, a}~{\cal K} ~( X_a, X_b; ~\Sigma_b - \Sigma_a ) ~\Psi ( x^\mu_a, \sigma^{\mu\nu}_a, \sigma^{\mu\nu\rho}_a..... ). \eqno (19)$$

Notice that the temporal evolution is recast in terms of the analog of the Fock-Schwinger-Feynman paramter 
$\lambda$ [18, 30] which depends on $\Sigma_{p+1}$ .  
For further details we refer to [18,26, 30 ]. 

\bigskip

\centerline{\bf 3 . Explicit Derivation of the Black Hole Entropy-Area Relations from the New Relativity Theory}

\bigskip

Having gone through a brief tour of the the postulates of the New Relativity, 
we shall present in this section  the simple steps 
to derive the Black-Hole Entropy-Area linear relation and its logarithmic ( and higher order ) corrections.
leaving all the details to references [20,21]. 
We only need to discuss in detail the generalized wave equations ( for the $p$-loop oscillator ) 
 in a Clifford-manifold that encode the 
dynamics such family of $p$-brane or $p$-loop harmonic oscillators .

The most simple expression ( all modes are frozen except the zero modes ) is to write the simplified wave equation 
for the $p$-loop harmonic oscillator , in units 
$\hbar=c=1$ :

$$\{ ~ - {1\over 2 \Lambda^{ p -1} }  [  {\partial ^2 \over \partial  x^\mu  \partial  x_\mu }
+ {\Lambda^2 }  {\partial ^2 \over \partial \sigma^{\mu\nu} 
\partial  \sigma_{\mu\nu}} + {\Lambda^4  } {\partial ^2 \over \partial \sigma^{\mu\nu\rho} 
\partial \sigma_{\mu\nu\rho} }   +......]  + $$
$${ m_{p+1} \over 2 L^2 }~   
[\Lambda^{2p} {x_\mu}^2 +  \Lambda^{2p-2} \sigma_{\mu\nu}^2 +......\Omega_{p+1}^2~]~\}~ \Psi =     
T ~ \Psi [x^\mu, \sigma^{\mu\nu}, \sigma^{\mu\nu\rho},.....  ]. \eqno (20)$$

The solutions of the latter $p$-loop oscillator wave equations where given in [21,22 ] in terms of 
the dimensionless ( rescaled ) variables :

$${\tilde x_\mu} = {\Lambda^p x_\mu \over L}.~~~{\tilde \sigma}_{\mu\nu} = 
{ \Lambda^{p-1} \sigma_{\mu\nu} \over L} .~~~~
 {\tilde \Omega}_{p+1} = { \Omega_{p+1} \over L}. \eqno (21)$$

where the analog of oscillator amplitude is $L$ given by : 

$$ L^2 = { \Lambda^{ p+1} \over m_{p+1} } \Rightarrow \Lambda^{ p+1 } <~ L~< ~ { 1\over   m_{p+1} }. \eqno (22)$$
where $m_{p+1} $ is the analog of the Compton momentum. 

$$\Psi \sim  exp~ [ - ~ (  {\tilde x_\mu}^2 +   {\tilde \sigma}_{\mu\nu}^2 +....                  ) ] 
~ H_{n_i} ( {\tilde x_\mu} )~ H_{n_{jk}} (  {\tilde \sigma}_{\mu\nu} )~ H_{n_{jkl}} (  {\tilde \sigma}_{\mu\nu\rho} )           .....\eqno (23)$$
we are expressing as usual the ground state as the Gaussian  and the excited ones by the product of the 
Hermite polynomials. The excitations of the $p$-loop oscillator are {\bf collective} ones given by the set of 
center of mass excitations; holographic area, holographic volume, .... excitations : 

$$ N = \{ n_i; n_{ij}; n_{ijk},......\}. \eqno (24) $$

The Tension is quantized as follows [ 21, 22   ] :

$$ T_N = ( N + {1\over 2} 2^D )m_{p+1 } \eqno (12) $$
the degree of the Clifford algebra in $D$ dimesnions ( `` number of  bits `` ) is  $ 2^D$. The $first~ 
collective$ excited state corresponds to setting all the quantum numbers equal to $1$ in 
full compliance with the principle of dimensional democracy ( $p$-brane democracy ) or poly-dimensional covariance  : 
{\bf all } dimensions must appear on the same footing :

$$\{ n_i = 1 ; n_{ij}= 1 ; n_{ijk}=1 ,......\} \Rightarrow N_1 = 2^D \eqno (25) $$

The degeneracy of the the ${\cal N}^{th} = k $  state is given as a function of $D, N$ :

$$dg~(D, N_k ) = { \Gamma ( 2^D + N_k ) \over \Gamma ( N_k +1 ) \Gamma (2^D )  }.~~~~
N_k = k~ 2^D.~~~ k = 1,2,3....  \eqno (26)  $$

The degeneracy of the {\bf first collective } excited state is $N_1  = 2^D $ and is naturally given by simply setting 
$N = 2^D $ in the above equation :

$$ dg~(N = 2^D ) = { \Gamma ( 2N ) \over \Gamma ( N +1 ) \Gamma (N) }  \eqno (27) $$

The Entropy is defined as the natural logarithm of the degeneracy . Hence taking the logarithm  and 
using Stirling's asymptotic expansion of the logarithms of the Gamma functions yields for the first collective 
excited state the following Entropy :

$$  Entropy = S = 2N ln(2) - { 1\over 2} ln(N) - {1\over 2} ln ( 4 \pi) - O(1/N). \eqno (28 ) $$

Before invoking  Shannon's information entropy by setting the number of holographic $p$-loop bits to  coincide precisely 
with the ratio of the $d-2$-dimensional area associated with a black hole horizon in $d$-dimensions ( $ d \not= D $ ) 
of radius $R$ and the area of radius $\Lambda$ 
in the following manner : $  N \sim A/G$ , one needs to justify this assumption. Most importantly is to answer the 
following questions :
\smallskip
 
Where does Einstein's  gravity come from ? How is it obtained as the long distance effective theory 
from the `'' gas ``  of highly-excited $p$-loop oscillators quanta  associated with 
New Relativty in {\bf C}-spaces ( Clifford manifolds ) ? 
\bigskip

${\bullet }$ At the moment we cannot answer such difficult questions ; however we 
recall that the low energy limit of string theory ( the effective string action after integrating out the massive string modes ) 
reproduces Einstein-Hilbert  action with the ordinary scalar curvature term plus a series of 
higher powers of the curvature. 
Einstein's gravity is recovered as the low energy limit from strings propagating in curved backgrounds.  
\smallskip
${\bullet}$ Fujikawa [23] has given strong reasons why Shannon's Information Entropy is related to the 
Quantum Statistical ( Thermodynamical)  Entropy. In particular in understanding the meaning of Temperature.
\smallskip 
${\bullet}$ Li and Yoneya [24] , among many others, have derived the 
Bekentein-Hawking entropy-area linear relation by taking the 
logarithm of the degeneracy of the highly excited massive ( super ) string states in $d$-dimensions. 
\smallskip

${\bullet}$ The New Relativity principle advocates that dimensions are in the eye of the beholder [1,2] . 
In one reference frame an observer sees a gas of $p$-loop oscillators in $ D$-dimensions in the 
first collective excited 
state of $ N = 2^D$. 
In another frame of reference another observer sees only strings in $ d $ dimensions (  $d\not=D$) 
in a very highly excited state $ n $.

If we identify Shannon's information entropy as the number of bits $ N = 2^D$ of 
the $p$-loop oscillator which is just the degree of the Clifford algebra in $D$ dimensions  , 
Shannon's information entropy can them be re-expressed in terms of the number of bits as follows : 

$$ N = S_{Shannon} = log_2 ~( 2^N ) = N = 2^D \Rightarrow Number ~of ~states = {\cal N} = 2^N = 2^{2^D}. \eqno (29) $$ 
Notice the {\bf double exponents} defining the number of states. The Black-Hole Horizon literarily is an $information$  horizon as well ! 

Now we are ready to make our {\bf only } assumption. We will identify the number of $p$-loop bits 
$ N = 2^D$ in $D$ dimensions to 
coincide precisely with the number of $area$ bits contained by a Black-Hole horizon ( in $ d$ dimensions ) 
of a given Area in units of the Planck scale. Namely it is the $ratio$ of the areas of radius $R$ and $\Lambda$ that gives the vlaue of the number of geometrical bits : 

$$ N = 2^D = { A_{d-2}( R)  \over A_{d-2} (\Lambda)  } \sim { A_{d-2} \over \Lambda^{ d -2 }  }. ~~~~ 
G_d  = \Lambda^{ d -2 }  \eqno (30)  $$
where we just wrote down the value of the Newton constant $G_d$ in $d$-dimensions as $ \Lambda^{d-2}     $ . 
The number of transverse dimensions to the radial $r$  and temporal coordinates $t$ 
 of a spherically-symmetric black hole is $ d - 2 $. So the Horizon area refers to a $ (d - 2) $-dimensional one. 

This is {\bf all}  we need to obtain precisely the Black-Hole entropy-area relation in the literature 
including the logarithmic corrections [22] , up to numerical coefficients , directly from eq-(28)   :

$$  S \sim [2ln(2)] ( A/G)  - { 1\over 2} ln(A/G) - {1\over 2} ln ( 4 \pi) - O(1/(A/G)) +....... \eqno (31) $$

To conclude : We have obtained in a very straightforward fashion not only the Bekenstein-Hawking 
entropy-area linear relation for a Black Hole in any dimensions  but also 
the logarithmic corrections plus higher order corrections [ 21, 22 ] . 
All of these corrections  appear with a minus sign 
and the entropy-area relation satisfies the second law of black hole thermodynamics  :  

$$ If~ A_3 > A_1 + A_2 \Rightarrow S(A_3) - S(A_1 ) - S(A_2 ) > 0. \eqno (32) $$ 
If two black holes of areas $A_1, A_2$ merge to give another black hole of area $ A_3 > A_1 + A_2$ 
then the resulting entropy {\bf cannot } decrease. 
For an upper bound on the values of $A_3$ see [ 21 ]. In Planck units we obtained : 

$$  N_1 N_2 > N_3 > N_1 + N_2 ~\Rightarrow ~ 
{ A_1 A_2 \over G^2 } > ~ { A_3 \over G }~ > ~ { A_1 + A_2 \over G }. \eqno (33)$$
For further details of all the technicalities behind this construction 
we refer to [21, 22 ] .

Perhaps the most salient feature is the intricate realtion between $d, D, R, \Lambda$ in all these relations. Based on the 
definition of number of geometrical bits as the $ratio$ of two areas, one for radius $R$ and the other for radius $\Lambda$ one can immediately infer :

$$ N = { A_{d -2} ( R ) \over A_{d -2} ( \Lambda )} = 2^D = ( { R\over \Lambda} )^{ d -2 } . \eqno (34)$$
Taking the logarithms on both sides we get the desired relationship among $ D, d, R, \Lambda $ : 

$$ D (ln 2) = ( d -2 ) ~ ln ({ R \over \Lambda}) \Rightarrow R = \Lambda \rightarrow d = \infty ~! . \eqno (35)$$

This is a remarkable conclusion. When $ R = \Lambda$ , for $ D \ge 2 $ ( dimensions where Clifford algebras are defined) 
one recovers autmatically Nottale's scale relativistic results that when one reaches the impassible Planck scale the 
fractal dimension of spacetime blow up.    
We can see also why $ N $ cannot be equal to unity. If $ N_1 = N_2 = 1 $ this $violates$ the second law of black 
hole thermodynamics since $ 1 + 1 = 2 > 1.1 $. One must have at $least$ $ N = 2 $ in order that 
$ 2 + 2 = 4 = 2.2 $ and the relation (33) is not violated. 

\bigskip

\centerline{ \bf 4. The Generalized Spacetime Uncertainty Relations }
\bigskip

\centerline{\bf 4.1. The String Uncertainty Relations follow from the New Relativity }

\bigskip

We have studied the $p$-loop harmonic oscillator using the $C$-space wave equations. The free case admits 
plane wave type solutions :

$$\Psi = e^{i~ (k_\mu x^\mu + k_{\mu\nu}\sigma^{\mu\nu} + k_{\mu\nu\rho}\sigma^{\mu\nu\rho}+....) }. \eqno (36)$$
Inserting this plane wave type of solution into the wave equation fro the free $p$-loop case yields the generalized 
dispersion relation :

$$ \hbar^2 ( k^2 + {1\over 2} \Lambda^2 (k_{\mu\nu})(k^{\mu\nu}) +  {1\over 3!} \Lambda^4 
(k_{\mu\nu\rho})(k^{\mu\nu\rho}) + .....) = {\Lambda^{2p} m_{p+1}^2 \over \hbar^{2p} } . \eqno (37) $$
This is just the extension of :

$$ p^2 = \hbar^2 k^2.~~~ p^2 = m^2 . \eqno (38)$$

On dimensional analysis and  using the principle of polydimensional covariance one can infer that :

$$ k^2 \equiv k_\mu k^\mu.~~~(k_{\mu\nu})(k^{\mu\nu}) = \beta k^4.~~~(k_{\mu\nu \rho})(k^{\mu\nu\rho}) = 
\beta k^6......\eqno (39)$$
where $\beta$ is a proportionality coefficient; i.e all dimensions are weighted by the same value in compliance with 
polydimensional covariance.  

Using this relation and inserting into the square root  of the dispersion relation one obtaines an $effective$ value of the Planck constant :

$$\hbar_{eff} = \hbar ( 1 + {1\over 4} \beta \Lambda^2 k^2 + O(\Lambda^4 k^4 )+....). \eqno (40)$$

Recurring to the well known relation ( due to the Schwartz inequality and that $|z| \ge |Im~z|$ ) : 

$$ \Delta x ~\Delta p \ge {1\over 2} |< [ {\hat x}, {\hat p } > | = {\hbar_{eff}  \over 2}. \eqno (41)$$
After a little algebra and using the relations :

$$\hbar k = p.~~~ < p^2 > ~\ge~ (\Delta p )^2.~~~<p^4 >~ \ge~ (\Delta p)^4....\eqno (42)$$
one arrives at :

$$ \Delta x ~\Delta p \ge {1\over 2} \hbar + {1\over 2} {\beta \Lambda^2 \over 4 \hbar} (\Delta p)^2 + O[( \Delta p )^4] 
 +... \eqno (43)$$

Finally one derives the string uncertainty relation by keeping the first two leading terms :

$$ \Delta x \ge {1\over 2} {\hbar \over \Delta p}  + {1\over 2} {\beta \Lambda^2 \over 4 \hbar} (\Delta p). \eqno (44)$$
Eq-(44) has for $minimum$ value of $\Delta x$ of the order of the Planck length $\Lambda$ corroborating once more that distances below the Planck scale have no physical meaning.

\bigskip

\centerline{\bf 4.2 The Full Blown Generalized Uncertainty Relations } 
\bigskip

The stringy uncertainty relations are $not$ the most fundamental ones. The contribution to an $infinite$ number of $p$-branes where $ p = 0, 1, 2, 3,....\infty$ to the effective Planck constant is : 

$$ \hbar^2_{eff} = \hbar^2~ \sum_{ r = 1}^{\infty} { (k \Lambda)^{ 2 (r - 1)} \over r ! } = { e^{ z^2} - 1 \over z^2} . ~~~
z \equiv k \Lambda . ~~~ k = || {\vec k}  || = || k_\mu k^\mu ||^{1/2}  \eqno (45)$$

Following the same procedure as above one recovers the $full$ blown uncertainty relations for Quantum Spacetime due to 
$all$ extended objects from $ p = 0$ all the way to $p = \infty$ : 

$$ \Delta x \ge \sqrt { 2} ~\Lambda ~{ e^{ (\Delta z)^2/4 } \over (\Delta z)^2 } ~
\sqrt { sinh [ { (\Delta z )^2   \over 2    }      ]        }. ~~~ \Delta z = (\Delta k ) \Lambda.  \eqno (46)$$
which yields a $minimum$ distance of $\Delta x \sim 1.2426 ~ \Lambda $; i.e compatible with Nottale's Scale 
relativity postulate that the Planck scale ( resolution ) is unattainable in Nature. It takes an $infinite$ amount of 
energy to resolve such scales ; i.e reaching infinite dimensions in the process.

The effective Planck constant is : 

$$\hbar_{eff} =  \hbar~  \sqrt 2 ~ { e^{ z^2 / 4 } \over z   } ~
\sqrt { sinh [ { (z )^2   \over 2    }      ]        }. ~~~ z \equiv k \Lambda. \eqno (47)$$
 
Which means that the $effective$ momentum-squared  of point particle moving in the bakground geometry of $C$-space is ( we do not include numerical factors for convenience ) : 

$$ p^2_{eff} ( p ) = k^2 ( \hbar_{eff} [k])^2 ~  \sim \hbar^2 k^2 +  \hbar^2 \beta k^4 \Lambda^2   +.. 
\sim (\hbar^2 k^2) +   (\hbar^2 k^2) (\beta \Lambda^2 k^2)  +..      
. \eqno (48)$$ 

Therefore the effective  $effective$ squared-momentum due to the total contribution of $all$ $p$-branes is :

$$  p^2_{eff}=  p^2  { e^{ (p\Lambda/\hbar  )^2/ 2  }  \over [(p\Lambda/\hbar )^2/2] } ~
sinh [ {(p \Lambda/\hbar  )^2 \over 2}  ]  .~~~ p = \hbar k. \eqno (49)$$

We would like to point out a common and widely spread $misconception$ about the modified uncertainty relations. 
One can perform a $noncanonical$ transformation from $(x, p) $ to a new pair of ( noncanonical ) variables $x', p'$   
such as : 

$$ Given~ [ x, p ] = i \hbar ~~~x \rightarrow x'.~~~ p \rightarrow p'.~~~  
[ x', p' ] = i \hbar_{eff} ( p' ) = [ x, p' ] = i \hbar { \partial p ' \over \partial p } . \eqno ( 50 ) $$

Hence to first order corrections the relationship between $p$ and $p'$ after setting the $\beta$ parameter to unity is : 

$$ p  = p( p' ) = \int  { d p ' \over h_{eff} ( p ' ) } 
\sim   \int_0^{p'}  { d p ' \over  [ 1 +  (p'\Lambda)^2/4\hbar^2  +...      ]     } ~ \Rightarrow  
p= {\hbar \over \Lambda} tan^{ -1} [ {p'\Lambda \over \hbar}      ] . \eqno (51)         $$
Inverting the last relation yields : 

$$ p' = p' ( p  ) =  {\hbar\over \Lambda}~ tan~[{ p \Lambda \over \hbar}]  . \eqno ( 52)    $$

We deem very important to emphasize that the new momentum $ p'$ of the ( noncanonical ) pair of new variables  $ x', p' $ 
( the commmutator of $x, p$ is $not$  preserved )  $must~ not$ be confused with the $effective$ momentum $p_{eff} ( p ) $ 
given by (48) . Eqs-(48-51 ) indicate  clearly that $p' \not=p_{eff}$ ! Upon setting the Planck momentum to be 
$p_{Planck} =\hbar/ \Lambda$ in eq-(52) one can see that in a sense one may $compactify$  the momentum simply 
by having the momentum 
values bounded as :  $0< ~(p/p_{Planck})~ < 2\pi$. And this justifies naturally the 
introduction of  an ultraviolet cuttoff . 
Based on the Ultraviolet/Infrared entanglement in 
Noncommutative Geometry one can also postulate 
an infared cuttoff, in addition to an ultaviolet cuttoff, consistent with Nottale's 
impassible upper length scale [5]. A sort of stringy $T$-duality analog.  This idea allow us to postulate a phase 
transition for the universe, from the metastable vacuum whose average dimension is close to $4 + \phi^3$,  
to the Noncommutative 
quasi-crystal phase of $dual$  dimension $ \phi^3 = [1/ (4 + \phi^3)] $ [9] .
This phase transition from the metastable vacuum to the final quasi-crystal phase of average dimensions equal to $\phi^3$ 
is completed once the size of the universe has reached the upper impassible $dual$ scale to the Planck scale in the fashion we shall indicate next.  
A rough   estimate of such upper scale ${\cal L}$ was given by the geometric 
mean relation between the Planck scale and the Hubble radius $ R_H \sim 10^{60} \Lambda $ : 

$$ R^2_H = \Lambda {\cal L }.~~~ ({R_H \over \Lambda } ) = ({\cal L \over R_H}) \Rightarrow 
\Lambda < ~R_H ~< {\cal L } \eqno (53) $$  
   
This sort of `` renormalization group `` argument was given by Nottale [5] and recently by us [9] 
which provides a very plausible and elegant resolution to the cosmological constant problem. The Universe self tunes 
itself along the renormalization group flow given by the scaling temporal evolution ( size ) as follows :

$$ { E_{vac} ( \Lambda ) \over E_{vac} ( R_H ) } = ({ R_H \over \Lambda })^2 \sim 10^{ 120 } .\eqno (54)$$

To finalize this section we wish to add that we have arrived  at similar results as Majid [12] : 
at scales close to the Planck scale we do $not$ have the standard Lorentz invariance but a 
Quantum Group deformation as indicated by Majid and others [12]. We will discuss the role of quantum groups, Braided 
Hopf Quantum Algebras, Braided QFT next in the construction of the Master Action Functional . 
Especially, the importance it has in order to recover, in the long distance limit , ordinary 
Einstein-Riemannian Geometry from the more fundamental $C$-space Geometry.

\bigskip

\centerline { \bf 5 Rigid Branes, Spin, Extrinsic Curvature and the Effective $C$-space Geometry } 

\bigskip

In this section we will provide with a $geometrical$ meaning to the effective momentum $p_{eff} $ 
appearing in eq-(48). To the author, this is probably one of the most fascinating results 
from the New Relativity . It is the emergence of an effective 
background geometry linked to a spinning particle and the actions with an explicit $extrinsic$  curvature 
( rigidity ) terms, directly from $C$-space.  
We shall argue why this effective Geometry may in fact be connected to ${\cal W} $ Geometry and Finsler Geometries [13] .
${\cal W} $ Geometry is the geometry related to higher conformal spin theories, from spin $1$ all the way to 
$\infty$.    
The subject of extended conformal field theories and ${\cal W} $ strings is very vast that we just refer to [34, 36] 
for references. 

Pavsic [30]  long ago has shown that the classical equations of motion for a $rigid$ 
$p$-brane in a $curved$ background can be derived from a Lagrangian which contains $extrinsic$ curvature terms . 
The world line for a rigid  point particle admits for equations of motion the 
Papapetrous' s equations ( also studied by Pezzaglia using Clifford algebras )
for a spinning particle ( not a geodesic ) . Though our rigid particle is $pointlike$ it has 
effectively $spin$ due to the $extrinsic~curvature$  term in the action which forces the
 particle to move in a $helical$ path. 

Pavsic was also able to show that the action for a rigid particle can be obtained via a 
`` Kaluza-Klein ``  reduction of an $open$ string wound up around a compact direction. 
The particle equations of motion follow from the $truncated$ 
string equations of motion. If the string is spacelike, and is compactified along a spacelike direction, 
the derived rigid particle action has only $tachyonic$ non-trivial solutions. 
This could be relevant to the current tachyon condensation studies in $M$ theory ( Sen and Ghoskal [8] ) and 
a plausible explanation of the apparent 
superluminal group velocities found in the recent experiments. 
It is clear from the generalized dispersion relation in $C$-space, 
given by eq-(37),  that one may encounter effective group velocities faster than light. This is not surprising since 
after all we are working in $C$-space and $not$ in the ordinary spacetime of Special Relativity. 
If the string is timelike then the derived 
rigid particle has subluminal ( bradyonic) non-trivial solutions corresponding to a helical ( or circular ) motion.
This helical motion is $precisely$ the one associated with the values of the effective $p_{eff}$ given by 
eq-(48), as we intend to show next.  

Pavsic considered the action for rigid $p$ branes in a target curved $D$ spacetime: 

$$ S = \int d^{p+1}  \sigma \sqrt { | \gamma |} ~( T - \mu g_{\mu\nu} H^\mu H_\mu ) . ~~~ 
\gamma_{AB}= \partial_A X^\mu \partial_B X^\nu g_{\mu\nu} . \eqno (55a)$$
where $\gamma_{AB}$ is the induced $p$-brane metric as a result of the embedding of the $p$-brane into spacetime. 
$\gamma$ is the determinant of $ \gamma_{AB}$. $ T$ is the $p$-brane tension; $\mu$ is the rigidity parameter ( like a `` friction `` term that opposes bending ) ; $X^\mu ( \sigma^A)$ are the embedding coordinates of the $p$-brane whose worldvolume coordinates are $\sigma^A$ with $ A = 1,2,3....p+1 < D$. 
$ H^\mu$ is a vector related to the $extrinsic$ curvature 
: 

$$ H^\mu_{AB} = D_A D_B X^\mu (\sigma^A)  + \Gamma^\mu_{\alpha \beta}~ \partial_A X^\alpha (\sigma^A)~ 
\partial_B X^\beta (\sigma^A) \equiv 
{\cal D}_A {\cal D}_B X^\mu (\sigma^A) . ~~~~ H^\mu \equiv H^\mu_{AB} \gamma^{AB} . \eqno ( 55b) $$
The Lagrangian is of $second ~order$ and the equations of motion contain quartic derivatives when $H^\mu \not=0$. 
When $H^\mu = 0$ the equations of motion coincide with the harmonic equations of motion associated with the 
minimal embedding surface in a curved background :
$ H^\mu = {\cal D}_A {\cal D}_B X^\mu = 0 $.  In 
the special case of a rigid particle the 
$fourth$ order equation associated with its worldline is nothing but Papapetrou's equation for a spinning particle ! :

$${1 \over \sqrt { \gamma} } {d p^\mu \over d \tau} + {1 \over \sqrt { \gamma} } \Gamma^\mu_{\alpha \beta}   
p^\beta {d x^\alpha \over d\tau} + {1 \over \sqrt { \gamma} } 
R_{\nu\alpha\beta}^\mu ~S^{\nu \alpha} 
{d x^\beta \over d\tau} = 0 . ~~~~ H^\mu { d X^\nu \over d \tau } = S^{\mu\nu}. \eqno (56)$$

This has a straighforward explanation based on $C$-space geometry. The $geodesic$ Clifford-algebra valued lines 
in $C$-space $are ~not$ 
geodesics in ordinary spacetime ! Secondly, to differentiate with respect to $\tau$ is $not$ the same than to 
differentiate with respect to the $\Sigma_{p+1}$ ( which is the true $C$-space generalization of proper time  ).  
The effective particle motion was shown to be $ helical$ and was due to the classical $spin$ , `` induced'' 
by the extrinsic curvature terms. The classical spin or  
intrinsic angular momentum ( associated with the $second$ order action ) is given by :

$$  S^{\mu\nu} = \pi^\mu {d x^\nu  \over d\tau}- \pi^\nu  {d x^\mu  \over d\tau}. \eqno (57)$$
where $ \pi^\mu$ is the $second ~order$ momentum conjugate to the variable $ {\dot X}^\mu$. Whereas the first order momentum $p^\mu$ is the conjugate to the variable $X^\mu$ and gives the standard orbital angular momentum 

$$  L^{\mu\nu} = p^\mu x^\nu - p^\nu  x^\mu.~~~ J^{\mu\nu} = L^{\mu\nu} + S^{\mu\nu}.             $$
The $second$  order action of Pavsic, involving extrinsic curvature rigidity terms, contains for basic variables $ X^\mu $ 
and $ ( dX^\mu / d\tau ) $ ( and their assocated first and second order canonical momenta ) .  This occurs also 
in the more fundamental Finsler Geometries ( Jet Bundles )   where the metric is 
both a function of the position and velocities. Finsler Geometries have both a maximum speed and maximum four acceleration
( maximum value of tidal forces ). The maximum four acceleration is $ a = c^2 / \Lambda$. If the Planck scale is seto to zero, Finsler Geometry collapses to Riemannian one. Finsler Geometry [13] has   natural connection to $W$ geometry ( see the references of C. Hull in [34]) .
And the latter is related to the $extrinsic$  geometry of embedded surfaces in $CP^N$ spaces as Sotkov, and then 
Gervais and Matsuo have shown, [36] . The relation to the Moyal-Fedosov 
Deformation Quantization was shown by the author following the work of C. Hull , see [34].  
 
All this has a natural interpretation in $C$-space ( Pezzaglia [2] has also discussed the relation between spin, Clifford algebras and polydimensional covariance ).  
One has an ordinary center of mass of motion associated with the $p$-loop histories. 
The first order corrections 
are the nothing but the $area$ remnants of the extended objects or $p$-loop holographic coordinates  : 
The holographic area proyections $\sigma_{\mu\nu}$  have for their  Fourier conjugates ,  
the two-vector $ k_{\mu\nu}$ .

By keeping only the leading term in the expansion , the effective momentum ( 48 ) 
will have two pieces, one from the center of mass motion, 
and another from the holographic components $k_{\mu\nu}$. 
The latter are the Fourier dual to the holographic Area coordinates $\sigma_{\mu\nu}$ or 
Spin tensor/Spin-two-vector $ S^{\mu\nu}$. 
The temporal change of the holographic area $\sigma_{\mu\nu}$/$S^{\mu\nu}$ is 
nothing but the $k_{\mu\nu}$ as we intend to show.

The effective momentum due to the center of mass and spinning motion found in [30] was :

$$p^\beta = [ m + \mu H^\mu H_\mu ] ~ [{ 1 \over \sqrt { \gamma}}  { dX^\beta  \over d\tau}]  
- { 1 \over \sqrt { \gamma}}  { dX_\alpha  \over d\tau} ~ 
[  p^\alpha  { 1 \over \sqrt { \gamma}}  { dX^\beta  \over d\tau} - 
p^\beta   { 1 \over \sqrt { \gamma}}  { dX^\alpha  \over d\tau} ]. \eqno (58) $$

To make this connection more explicit with the effective $p_{eff} $ found in eq-( 48 ) 
we need firstly to study the special solutions  which correspond to the case when 
$H^\mu \not=0$ but with $ H^\mu H_\mu = 0$ ( null-like acceleration ); i.e  
these solutions precisely correspond  to the helical motion in flat spacetime.  
Notice that in general flatness 
does not necessarily imply  torsionless. Flat Superspace Supergravity $has~ Torsion$. The  
torsion is associated  to $spin$. The connection in superspace has the ordinary Levi-Civita connection but an extra piece due 
to the fermion bilinear terms ( torsion ) . In this special case the rigid point particle follows a 
$helical$ world line ( as a result of the spin )  
with center of mass momentum given by the $translational$ motion along the axis of the helix and 
the angular rotation frecuency is precisely related to the rigidity parameter $ \omega^2 = (m/2\mu)$. 
The vector $ H^\mu $ satisfied $ H^\mu = (d^2 X^\mu / d\tau^2)$ ( an acceleration ) and it had a $null$ norm 
$H^\mu H_\mu = 0$. 
The rate of change of the Spin two-vector ( Areal coordinates  ) is :

$$ { {\cal D} S^{\alpha \beta} \over {\cal D} \tau } = - 
[  p^\alpha  { 1 \over \sqrt { \gamma}}  { dX^\beta  \over d\tau} - 
p^\beta   { 1 \over \sqrt { \gamma}}  { dX^\alpha  \over d\tau} ] . \eqno (59a)$$
Which clearly has the form of a two-vector momentum : 

$$ p^{\alpha\beta} = m^2 \Lambda^2 k^{\alpha\beta} 
 \leftrightarrow  m  [  p^\alpha  { 1 \over \sqrt { \gamma}}  
{ dX^\beta  \over d \tau} -  p^\beta { 1 \over \sqrt { \gamma}}  { dX^\alpha  \over d \tau} ] .~~~ 
\lambda = ( {\Sigma_{p+1} \over m_{p+1}} )^{ (1/2p+2 )} ~\leftrightarrow {1 \over m }   \eqno (59b)$$
where $\lambda$ is the the analog of the Fock-Schwinger-Feynman evolution parameter ( a length scale )  
induced in eq-(11).

In the proper time gauge $ \gamma = 1 $, when $H^\mu H_\mu = 0$, the rigidity constant $\mu$ 
$decouples$  from eq-(58 ) ( not from the theory )  : 

$$p^\beta = m { dX^\beta  \over d \tau}   
- { dX_\alpha  \over d\tau} ~ 
[ p^\alpha  { dX^\beta  \over \tau} - 
 p^\beta   { dX^\alpha  \over d \tau} ]. \eqno (60) $$

one can compare then : 

$$p^\beta  = m { dX^\beta  \over d \tau}  - { dX_\alpha  \over d \tau} ~ [ p^\alpha  { dX^\beta  \over d \tau} - 
 p^\beta  { dX^\alpha  \over d \tau} ] ~ \leftrightarrow ~p_{eff} = 
(\hbar k^\beta )  +  ( \hbar k_\alpha  k^{\alpha\beta } \Lambda^2  ). \eqno (61)$$ 
Hence : 

$$  m { dX^\beta  \over d \tau} \leftrightarrow \hbar k^\beta .~~~ 
m {dX_\alpha  \over d \tau} ~ [ {p^\alpha \over m }   { dX^\beta  \over d \tau} - { p^\beta \over m }  { dX^\alpha  \over d \tau} ]
~\leftrightarrow~ \hbar k_\alpha ~ k^{\alpha\beta } \Lambda^2 . ~~~ m \leftrightarrow  { 1\over \lambda}. \eqno (62)$$
Which means then : 

$$ m~[ {p^\alpha }   { dX^\beta  \over d \tau} - { p^\beta }  { dX^\alpha  \over d \tau} ]
~\leftrightarrow~   m^2 k^{\alpha\beta } \Lambda^2 = p^{\alpha\beta} . \eqno (63 )$$

We have not finish yet. We still need to be more precise. If we want to match exactly the quantity :
$ \hbar^2 k^2 \beta \Lambda^2 k^2 $ appearing in the second term of the expression for $p^2_{eff}$ in 
eq-(48)  with the square of the last term of eq-(61) :  
$ (\hbar k_\alpha  k^{\alpha\beta}\Lambda^2 )(\hbar k^\alpha  k_{\alpha\beta} \Lambda^2)$  
we $must$  have a proportionality factor $(m^2\lambda^2_h)$ between them as follows. 

Using the condition that $k^{\alpha\beta} k_{\alpha\beta} = \beta k^4 $ and the relation : 
$k_\alpha  k_\beta k^{\alpha \beta } = 0$ due to the $antisymmetry$ of the two-vector $k^{\alpha\beta} $ one arrives :  
 $$ (\hbar^2 k^2)( \beta \Lambda^2 k^2) = 
( m^2 \lambda^2_h ) (\hbar^2 k^2) ( k^{\alpha\beta} k_{\alpha\beta}) (\Lambda^4) = 
( m^2 \lambda^2_h ) (\hbar^2 k^2) (\beta k^4)  \Lambda^4 ~ 
\Rightarrow  ~ (m^2) (\lambda^2_h) (k^2) (\Lambda^2) = 1 . \eqno (64) $$

This last equation (64) in conjunction with the mass-shell condition forces a relationship among  
$\Lambda, ~\lambda_h ,~ m $ 
as follows : 
$$Given~~~ m^2 = k^2 ~\Rightarrow {1 \over m^2} = (\lambda_h) (\Lambda ) ~\Rightarrow  \Lambda < { 1\over m } < \lambda_h  
. \eqno (65a)$$
And $once$ again we recover a $geometric$ mean relation among $3$ scales : $\Lambda; \lambda_h; (1/m ) $ as we did 
for the $p$-loop harmonic oscillator earlier on in the derivation of the Black-Hole Area-Entropy 
relation and its logarithmic corrections. The physical meaning of the $geometric$ mean relation is the following :

The rigid particle follows a $helical$ world line in such a fashion [30] that the circular motion moves along a 
$null$ surface in such a way the the net spacetime interval  spanned by its motion is precisely equal to the 
spacetime interval spanned by the center-of-mass motion ( see [30] for details ). If one defines the temporal-step  
interval parameter to complete one full revolution to be $ \lambda_{helix} \equiv  \lambda_h$ then 
the geometric mean relation is clear. The  
temporal-step parameter, or period,  
to complete one full revolution is greater or equal to the Compton wavelength of the particle 
( in units of $c=1$). 
This immediately allows us to evaluate the value of the $rigidity$ parameter $\mu$ from the frecuency relation [30] :

$$ \omega^2 = {m \over 2\mu} = ( {2\pi \over \lambda_h}  )^2 ~\Rightarrow \mu = 
{ m \lambda^2_h \over 2 ( 2\pi)^2 } . 
\eqno (65b) $$  
From eq-(65a) one can deduce the expression for the rigidity parameter that has dimensions of length : 

$$ \mu = { 1 \over 2 ( 2\pi )^2 m^3 \Lambda^2 } = 
{ 1\over m} { 1 \over 2 ( 2\pi )^2} ({ 1\over m \Lambda})^2. \eqno(65c)$$
If one sets the value of $m$ to be of the same order of the Planck mass $M_p \sim { 1/\Lambda } $ 
one gets a rigidity value of :

$$ \mu = { 1 \over 2 ( 2\pi )^2 m^3 \Lambda^2 } = \Lambda { 1 \over 2 ( 2\pi )^2} . \eqno(65d )$$
which will give us a natural estimate for the value of the rigidity parameter in Nature. 
It would be interesting to explore other values for $\mu$ and see if there are any Astrophysical signals [30] of 
its existence.     

The most important conclusion of this section is that a geodesic in $C$-space does  
$not$ correspond to a geodesic in ordinary spacetime. The first leading corrections of the effective $C$-space Geometry 
furnish naturally the Papapetrou's equations for a rigid particle . ; i.e a classical spin is $induced$ 
by the extrinsic curvature rigidity terms   
and, consequentlty,  the  worldline is a $helical$ ( circular in some limiting case ) motion in ordinary spacetime; 
i.e the rigid particle 
does $not$ follow  a geodesic, the four acceleration is $not$ zero, although it has null norm .  
Roughly speaking, the free ``multidimensional- particle `` in $C$-space, to the first leading approximation, 
corresponds to a spinning particle ( helical motion ). 

We took the Planck scale $\Lambda$ to be the natural length scale of this problem because of its 
relation to the $C$-space geometry. There are also the two other natural length scales  $\lambda_h$ 
( the helical motion period ) and $1/m$  ( the Compton wavelength ) 
 and we were able to derive the $geometric$ mean relationship among these  $3$ scales.  
The latter relation allowed us to estimate the value of the rigidity constant $\mu$ exactly in terms of $ m, \Lambda$. 
The Compton wavelength of the rigid particle falls in between the Planck scale and the $\lambda_h$ scale ( period )  
which emerged from the full $C$-space propagator. 
Pezzaglia [2] took another  different value for the natural scale related to the properties of the electron.  
The relevant result is the $emergence$ of classical spin encoded in the Clifford-algebraic construction of 
$C$-space. For interesting ideas pertaining the role of Grassmanian time in QM see [39] and for superluminal travel through extra dimensions see [40].  

\bigskip

 \centerline{\bf 6.   The Four Dimensional Conformal Anomaly, Fractal Spacetime and the Fine Structure Constant }
\bigskip

In this section we will briefly summarize the most relevant features of Fractal spacetime to cosmological applications. 
In essence, the universe began as a process of non-equilibrium self-organized critical phenomena. 
Following the spirit of Cantorian-Fractal spacetime [6] in [9] we 
computed the effective ( time dependent ) average dimension of the world by taking the 
statistical average of an infinity 
family of $p$-loops , of $bubbles$ whose dimension ranged from $ D = - 2$ to $D = \infty$. The gamma function was derived 
as the dimensional ensemble distribution. The average dimension ( observed today ) $relative$  to the $zero$ point dimension $ D_o = -2 $ was : 

$$ < D - D_o > =   < D' > = { \int^{\infty}_0 ~ dD'  ~ D' ~{\sqrt \pi}^{D'} ~[  \Gamma ( { D' + 2 \over 2 } ) ]^{-1} 
\over  \int^{\infty}_0  ~ dD' ~{\sqrt \pi}^{D'} ~[  \Gamma ( { D' + 2 \over 2 } ) ]^{-1}     } \sim 6.236 
$$ 
$$\Rightarrow ~< D > =  [6.236... + ( - 2 )]  = 4.236 = 4 + \phi^3 . \eqno (66)$$
where $\phi$ is the Golden Mean : $ { ({\sqrt 5} - 1)/2 } = 0.618...$.  
The average dimension was of the same magnitude as the average 
dimension of the transfinite Cantorian-Fractal spacetime ${\cal E}^{ (\infty)}$ developed by 
M. S. El Naschie [6]. The average dimension  $< ~{\cal E}^{ (\infty)    } ~> ~ = 4 + \phi^3 $ 
coincides precisely with the 
Hausdorf dimension of the set $ dim~{\cal E}^{ ( 4 ) } = 4 + \phi^3 = {1 / \phi^3 } = ( 1 + \phi)^3 $ 
that is packed densely onto a smooth manifold of four topological dimensions. 
Using the $bijection$ formula [6] : 

$$ dim~ {\cal E}^{ ( n ) } = (  { 1\over \phi}  )^{ n -1 } ~\Rightarrow dim~ {\cal E}^{ ( -2  ) } = 
\phi^3 = { 1 \over 4 + \phi^3 } = { 1 \over dim ~  {\cal E}^{ ( 4 ) } }  . \eqno (67)$$
Hence the set  ${\cal E}^{ ( -2  ) }$ ( Hausdorff dimension equal to $\phi^3$ and densely 
embedded into  a smooth space of topological dimension $ - 2 $ ) is the dimension $dual $ to the set 
 ${\cal E}^{ ( 4 ) }$  ( Hausdorff dimension equal to $4+ \phi^3$ and densely 
embedded into  a smooth space of topological dimension $ 4  $ ).                             .    

The backbone set ${\cal E}^{ (0) } $ 
is packed densely onto a point , topological dimension $0$, and 
its  Hausdorff dimension equals the Golden Mean $\phi$. The backbone set is 
a random Cantor set whose dimension is $\phi < ln (2 ) / ln (3 ) $ with probability $one$  
according to the celebrated Maudlin-Williams theorem.   
The transfinite Cantorian-Fractal spacetime model of El Naschie is the $randomly ~constructed $ geometric space that 
fits  very naturally within the Random Process Physics  ( Self Referential Noise as a model of reality ) 
program of Cahill and Klinger [29] ,  based on Godel and Chaitin's work 
, and consistent with the most recent work on Quantum Information Theory and 
 Quantum Algorithmic Random Processes  in Nature [31] .     
In addition, it adopts von Neumann's Noncommutative Pointless Geometry at its very core. 

The authors [32] more than two years ago computed the intrinsic Hausdorff dimension of spacetime at the 
$infrared$ fixed point of the quantum conformal factor in $4D$ Gravity.  
The fractal dimension was determined by the coefficient $Q^2$ of the 
Gauss-Bonnet topological term associated with the four dimensional conformal anomaly ( trace anomaly ) and was computed to 
be $greater$ than four.  We were able to show that one can relate the value of 
the Hausdorff dimension computed by [32] 
to the universal dimensional fluctuation of spacetime given by $ \phi^3 / 2 = 0.11856...$ [38].  
Based on the infared scaling limit and using recent Renormalization group arguments by El Naschie [6] 
we conjectured that the $unknown$ coefficient $Q^2$ 
associated with the four dimensional conformal anomaly may be equal to the inverse fine structure constant of values ranging from $137.036$ to $ 137.641$ [38] . The idea is based on Eddington's old belief that the inverse fine structure constnat could play the role of an $internal$ ( electron's dimension ).

The Hausdorff dimension computed in terms of the conformal anomaly coefficient $Q^2$ by [32] was : 

$$ d_H = 4~ { 1 + \sqrt { 1 + {8 \over Q^2 }} \over  1 + \sqrt { 1 - {8 \over Q^2 }} } ~\ge ~ 4 . \eqno (68) $$
Inverting this relation allows one to express directly the anomaly coefficient $Q^2 $ in terms of $d_H$. 
By re-writing : 

$$ d_H = 4 + \epsilon = 4 ( 1 + \epsilon /4 ) = 4 ( 1 + \phi^3/8) = 4 \delta ~ \Rightarrow 
Q^2 = 2~ { (\delta^2 + 1 )^2 \over \delta ( \delta^2 - 1 ) } = 137.6414382326 . \eqno (69)$$
Thus, if one sets $ 4 + \epsilon $ to be $4$ plus the universal dimensional fluctuation $\epsilon = \phi^3/2 $ one 
obtains for the anomaly coefficient $Q^2$ a value very close to the experimental value of $137.036$. The value of 
$137.641..$ will be then associated with the $infrared$ contribution to the inverse fine 
structure constant due to the quantum fluctuations  of the conformal mode of the metric ! 

\bigskip

\centerline{\bf 7.   Noncommutative Geometry, Negative Probabilities and Cantorian Fractal Spacetime  }

\bigskip

Recently [38] we provided a straightforward explanation of the Young's double-slit experiment of a QM 
particle based on the Noncommutative Geometric nature of the transfinite spacetime 
${\cal E}^{(\infty)  }$ and Negative probabilities. Since spacetime is in essence a $randomly$ constructed space and the 
path of the QM particle is $fractal$ then any question about the exact spatial location of a microscopic point is 
fundamentally $undecidable$ due to the inherent uncertainty and fuzziness of the geometrical structure of 
such space [6]. Since von Neumann's Noncommutative Geometry is a $pointless$ one a `` point `` in  ${\cal E}^{(\infty)  }$ can in fact occupy $two$ different locations at the same time; i.e unions and intersections are indistinguishable. The same argument follows from Pitkannen's construction of $p$-Adic Fractals : there is a fundamental indeterminism in these spaces. The Topology of ${\cal E}^{(\infty)  }$ is in fact a $p$-Adic topology : every point is the center of a disk because every ``point''  can occupy many places at once.

The Young's double-slit experiment admitted a straightforward interpretation by simply assigning to the particle a 
fractal zig-zag motion around the two slits $A, B$ : a Peano-Hilbert curve. The negative probability came very natural due to the $opposite$ orientations of the two Peano-Hilbert curves around slit $A, B$ . 
The probability was related to the $inverse$ fractal dimensions of the sets around slit $A, B$ . 
The respective probabilty assignments were : 

$$p(A) = \phi.~~~p(B) = - \phi^2. ~~~p( A \wedge B ) = - ~ p(B \wedge A) = - \phi^3. ~~~ p(A|B ) = \phi. ~~~ 
p(B|A) = \phi^2 . \eqno (70 )$$
where the last three expressions are the joint probability for the simultaneous coexistence of the particle
at $A$ and $B$ , and the conditional probabilities, in such a way that the total sum of probabilities yields $unity$ exactly :

$$ [\phi - \phi^2 - \phi^3 ] + [ \phi + \phi^2 ] = 0 + 1 = 1  . \eqno (71)$$
Since 

$$\phi + \phi^2 = \phi ( 1 + \phi ) = 1.~~~ \phi - \phi^2 = \phi ( 1 - \phi) = \phi \phi^2 = \phi^3. ~~~ 
\phi = { 1\over 1 + \phi} . \eqno (72)$$

\centerline { \bf 8. Quantum Groups, Braided Hopf Quantum Clifford Algebras and the Master Action Functional }  
\bigskip 

It was argued by the author in [1] that the master action functional for the nested family of $p$-loop histories could be given by a Braided QFT given by the action :

$$ S = \int [D X (\Sigma)] ~ {1\over 2} ~ 
{ \delta \Psi [ ( X (\Sigma) ] \over \delta X ( \Sigma)~}  *~ {\delta \Psi [ ( X (\Sigma) ] \over \delta X ( \Sigma) } 
+  ({m_{p+1}})^2 ~\Psi [ ( X (\Sigma)] ~*~ \Psi [ ( X (\Sigma)]   +  $$
$$ {g_3 \over 3 !} ~\Psi [ ( X (\Sigma)] ~*~ \Psi [ ( X (\Sigma)]~*~ \Psi [ ( X (\Sigma) ] +  
+ {g_4 \over 4 !} ~\Psi [ ( X (\Sigma) ] ~*~ \Psi [ ( X (\Sigma)] ~*~\Psi [ ( X (\Sigma)] ~*~  \Psi [ ( X (\Sigma)]. \eqno (73 )$$

Where the action functional is invariant under a Braided Hopf Quantum Clifford algebra associated with the 
Quantum-Clifford-algebra valued master field $\Psi [ ( X (\Sigma) ]$. 
The $ X ( \Sigma)$ are the Clifford algebra valued hyperlines ( higher dimensional version of twistors for example or hypercomplex numbers )  
in $C$-space parametrized by the $C$-space extension of the `` proper time `` $\Sigma$ as shown in eq-(4) . 

As mentioned earlier, the first  term  corresponds to the quadratic kinetic term. 
The second ones to the analog of mass-squared terms for a scalar 
field theory ; the triple interaction vertex corresponds to both a product ( two hyperlines join to give a third one ) and co-product ( one hyperline breaks into two ) of the Quantum algebra.  
The quartic terms correspond to the braided scattering of the four lines. These  are the 
only terms allowed consitent with the Noncommutative Braided Hopf Quantum Clifford algebra [15] encoded by the 
star product $*$.

This star product is reminiscent of the Noncommutative star  product associated with the 
$BRST$ and Batalin-Vilkoviski formulation of string field theory [16, 17] where anticommuting variables are introduced. 
The string field $\Psi [ X^\mu , c ] $ is now a functional of the string $world-sheet$ coordinates 
$X^\mu ( \sigma^a)$ and of the 
ghost coordinates $c (\sigma^a )$ with $ a = 1, 2 $. Compare this with the  
Master field which is a Clifford-line functional. The 
$C$-space Clifford-algebra valued ( matrix ) multivector $X ( \Sigma_{p+1} ) $ that solely depends on one parameter.

Similar deformed star products have been obtained by Vasiliev in his construction of higher 
conformal spin algebras from the deformation of the Anti de Sitter Algebra. These 
higher spin theories have been essential to construct Higher Spin Supergravity Theories which are conjectured to be the effective field theory limit of $M$ Theory compactifications on $ S^7 \times AdS_4$ [38]. 

This Braided QFT [33] is highly nontrivial. To derive the wave equations used for the $p$-loop harmonic oscillator we 
had  to make several assumptions. The first one is to freeze or quench the higher order modes of the 
functionals so one can 
approximate the dynamics by writing $ordinary$ differential equations. Secondly one must assume  a $flat$ $C$-space metric.
and thirdly we had to set the cubic and quartic couplings to zero . See [20, 21]. In general one has a $nonlinear$ oscillator which has far richer properties than the linear case.

\bigskip

\centerline { \bf 9. Future Prospects, Riemannian Geometry as the large distance limit of $C$-space Geometry }   

Majid has proposed a Planck scale Hopf algebra, a sort of gravity/co-gravity duality : meaning a 
$curved ~phase$ space ( perfectly consistent with the Finsler and ${\cal W}$ Geometry )  
to formulate the quantum group algebraic properties of a plausible Planck scale quantum geometry 
that should reproduce ordinary classical Riemannian Geometry in the long distance limit. 
In particular he derived the analogous ( not identical ) commutations relations to our commutation relations eq-(41) :

$$ [ x , p ] = i\hbar_{eff} = i\hbar [ 1 - e^{ - x /L } ] . \eqno(74) $$
where $L$ is a suitable length scale. The quantum flat space limit is attained when $ L = 0$.   
The classical limit is attained when $\hbar = 0$. The most interesting aspect of 
Majid's  Planck scale Hopf Algebra construction is that one can simulate the dynamics of a particle with a $position$ dependent momentum $p'$ given by :

$$ p' = p_o[ 1 - e^{ - x /L } ] .~~~ 
\Rightarrow [ x', p' ] = [ x, p_o( 1 - e^{ - x /L })]=i \hbar_{eff}. \eqno(75)$$
In such a fashion that it $mimics$ the motion of a free falling particle into a black hole whose velocity 
measured by an asymptotic observer at infinity is :

$$ v( r ) = v_{\infty} ( 1 -   { 1 \over 1 + r/L +...} ) + O(\hbar ) . \eqno ( 76) $$
As the particle approaches the origin it appears to move more slowly relative to the asymptotic observer 
and will take an infinite amount of time ( relative to the observer ) to reach the origin. This is 
exactly similar 
( up to a factor of $1/2$ ) to the formula for 
the radial infalling particle velocity in the vecinity of a Black Hole of mass $M$ and Schwarschild radius $R$ : 

$$ v( r ) = v_{\infty} ( 1 -   { 1 \over 1 + r/2L } ).~~~ 
L = { G M } = { 1\over 2} ( 2GM ) = {1\over 2 } R_{Schwarzchild}. ~~~ c = 1\eqno (77a)$$  

Another way to recapture the Riemannian ( Black Hole ) Geometry in the large distance limit $ x = r >> L $ is 
to consider using the full $p$-loop  extension of the 
Wheeler-de Witt equation  and find its solutions $\Psi$. 
This $p$-loop  extension of the Wheeler-de Witt equation could be simplified by 
using some sort of Minisuperspace approximation schemes and be reduced to an 
extension of the well studied Nonlinear Schroedinger equations and its solitonic solutions; i.e  
Nonlinear wave mechanics. Then one could look at the first linearize approximation 
and evaluate the expectation values of the multicomponents of the $C$-space Geometry metric components :
$  G_{\mu\nu} , G_{\mu\nu\rho\tau}........ $ in the large excitation 
level $n$ limit and show that :

$$ < \Psi | G_{\mu\nu} [ X ] | \Psi > =  
< \Psi | G_{\mu\nu} [x_\mu; \sigma_{\mu\nu}; \sigma_{\mu\nu\rho},... ] | \Psi > ~ \rightarrow  
g_{\mu\nu } ( Schwarzschild) + O( r/\Lambda)..... \eqno ( 77b ) $$

  $$ < \Psi | G_{\mu\nu\rho\tau} [ X ] | \Psi > =  
< \Psi | G_{\mu\nu\rho\tau} [x_\mu; \sigma_{\mu\nu}; \sigma_{\mu\nu\rho},... ] | \Psi > ~ 
\rightarrow 0.~~~ etc ..... \eqno ( 79c ) $$
where the $\Psi = \Psi [ G_{\mu\nu}, G_{\mu\nu\rho\tau}....]$ is a solution of the $p$-loop extension of the 
 Wheeler-de Witt equation . This is currently under investigation. Essentially one is looking at the geometry 
induced by a self-graviating gas of $p$-loops in the large distance limit. Gas which will condense and reproduce the Black Hole Geometry in the long distance limit.

Concluding : 

Based on what we have shown in this work it seems undoubtedly that the New Relativity is marching forward and that it will embrace many branches of modern Physics and Mathematics.

\smallskip

\centerline{\bf Acknowledgements}
\smallskip 

We thank A. Granik, C. Handy, E.Spallucci, S.Ansoldi, E. Gozzi, T. Smith, G. Bekkum, D. Finkelstein, 
A. Schoeller, J. Boedo, S. Duplij, M.  Pavsic, E. Guendelman, G. Kalberman, L. Nottale , M.S. El Naschie , 
D. Chakalov, M. Pitkannen, W. Pezzaglia,  S. Paul King, L. Baquero,  J. Mahecha, J. Giraldo for many discussions.  

Special thanks go to B. G. Sidharth for his kind invitation to Hyderabad, India. 
\bigskip

\centerline{\bf References}
\bigskip

1- C. Castro : " Hints of a New Relativity Principle from $p$-brane Quantum Mechanics " hep-th/9912113.

Journal of Chaos, Solitons and Fractals {\bf 11} (11) (2000) 1721.

C. Castro : " The Search for the Origins of $M$ Theory : Loop QM, Loops/Strings  and Bulk/Boundary Duality " 
hep-th/9809102.

C. Castro : `` $p$-Brane Quantum Mechanical Wave Equations  `` .hep-th/9812189

2. W. Pezzaglia : " Dimensionally Democratic Calculus and Principles of Polydimensional Physics " gr-qc/9912025.
To appaer in the Int. Jour. Theor. Physics.

3- C. Castro : " Is Quantum Spacetime Infinite Dimensional ?" hep-th/0001134. 

Journal of Chaos, Solitons and Fractals {\bf 11} (11) (2000) 1663- 

4- C. Castro, A. Granik  : " How the New Scale Relativity resolves some Quantum Paradoxes  " 

Journal of Chaos, Solitons and Fractals {\bf 11} (11) (2000) 2167.   

5. L. Nottale : Fractal Spacetime and Microphysics, Towards the Theory of Scale Relativity

World Scientific 1992.

L. Nottale : La Relativite dans Tous ses Etats. Hachette Literature. Paris. 1999.

6. M. El Naschie : Jour. Chaos, Solitons and Fractals {\bf vol 10} nos. (2-3)  (1999) 567.

M. El Naschie : Jour. Chaos, Solitons and Fractals {\bf vol 9 } nos. (3)  (1998) 517.

M. El Naschie : Jour. Chaos, Solitons and Fractals {\bf vol 12 } nos. (2001)   179 .

M. El Naschie : Jour. Chaos, Solitons and Fractals {\bf vol 8 } no. ( 11 ) (1997)  1873  .

7-. G. Ord  : Jour. Chaos, Solitons and Fractals {\bf vol 10} nos. (2-3) (1999) 499.  

8- M. Pitkannen : `` Topological Geometrydynamics `` Book on line  http ://www. rock. helsinki.fi/$\sim$ matpitka 
   
M. Pitkannen : `` $p$-Adic TGD : Mathematical Ideas " hep-th/9506097. 

A. Khrennikov : " $p$-Adic numbers in Classical and QM ...."  quant-ph/0003016.

M. Altaisky, B. Sidharth : Jour. Chaos, Solitons and Fractals {\bf 10} (2-3) (1999) 167.  

V. Vladimorov, I. Volovich, E. Zelenov : " $p$-Adic Numbers in Mathematical Physics. World Scientific, Singapore 1994. 

L. Brekke, P. Freund : Physics Reports {\bf 231} (1993) 1-66.  

D. Ghoshal, A. Sen : " Tachyon Condensation and Brane Descent Relations in $p$-Adic String Theory " hep-th/0003278. 
 
9- C. Castro, A. Granik, M. S. El Naschie : " Why we live in $3+1$ Dimensions " 
hep-th/0004152.

10. C. Castro : Foundations of Physics Letts {\bf 10} (1997) 273.

C. Castro : `` the String Uncertainty Relations follow from the New Relativity Principle `` 
 Foundations of Physics {\bf 30} ( 8 ) ( 2000 ) 1301.  
   
11. A. Connes : Noncommutative Geometry. Academic Press. New York. 1994.

12.  S. Majid : Foundations of Quantum Group Theory. Cambridge University

Press. 1995. Int. Jour. Mod. Phys {\bf A 5} (1990) 4689.

S. Majid : `` Meaning of Noncommutative Geometry and the Planck Scale Quantum Group `` hep-th/0006166

A. Connes, D. Kreimer : `` Renormalization in QFT and the Riemann-Hilbert problem ...'' 
hep-th/0003188 . 

A. Connes, D. Kreimer : Comm. Math. Phys {\bf 210} ( 2000 ) 249.

L.C. Biedenharn, M. A. Lohe    :  Quantum Groups and q-Tensor Algebras . 
World Scientific. Singapore . 1995.

13- H. Brandt : Jour. Chaos, Solitons and Fractals {\bf 10} nos 2-3 
(1999) 267.

14- S. Adler : Quaternionic Quantum Mechanics and Quantum Fields .
Oxford, New York. 1995.

15-. Z. Osiewicz " Clifford-Hopf Algebra and Bi-Universal Hopf Algebra " q-alg/9709016. 

16-. E. Witten : Nuc. Phys. {\bf B 268} (1986) 253.

17-. B. Zwiebach : Nuc. Phys. {\bf B 390} (1993) 33. 

18-S. Ansoldi, C. Castro, E. Spallucci : Class. Quant. Gravity {\bf 16} (1999) 1833. hep-th/9809182. 
S. Ansoldi, A. Aurilia, C. Castro, E. Spallucci : `` Quenched-Minisuperspace Bosonic $p$-Brane 
Propgator `` University of Trieste preprint. To be submitted tp Physical Review {\bf D}. 

19--L. Smolin, S. Kaufmann : " Combinatorial Dynamics in Quantum Gravity " hep-th/9809161. 

20-C. Castro, A. Granik : `` Derivation of the Logarithmic Corrections to the 
Black-Hole Entropy from the New Relativity `` submitted to the International Journal 
of Theoretical Physics. 

21-C. Castro, A.  Granik : `` P-loop Oscillator on Clifford manifolds and Black Hole Entropy `` 

physics/0008222 v2.

22-P. Majumdar : ``Quantum Aspects of Black Hole Entropy `` hep-th/0009008.

23-K. Fujikawa `` Shannon's Statistical Entropy and the {\bf H}-Theorem in 
Quantum Statistical Mechanics `` cond-mat/0005496 

24- M. Li, T. Yoneya : Journal of Chaos, Solitons and Fractals `` {\bf 10} (2-3 ) ( 1999 ) page 429. 

25. P. Lounesto : `` Clifford Algebras and Applications `` Lecture Notes in Mathematics

26. C. Castro, M. Pavsic : `` The Multidimensional-Particle Propagator from the New Relativity `` 
CTSPS-preprint ( Atlanta ) and IJS-Ljubljana preprint ( 2000 ) .  To appear. 

27-B.G. Sidharth : `` Fractal Statistics `` physics/0009083 

B.G. Sidharth : `` Quantum Superstrings and the Quantized Fractal Spacetime `` physics/0010026 

W. da Cruz : `` Fractal Statistics, Fractal Index and Fractons : `` hep-th/0008222''  

28- H. Maris : Jour. Low Temperature Physics {\bf vol 120} ( 2000 ) 173.

29-R. Cahill , C. Klinger, K. Kitto  : `` Process Physics : Modelling Reality as Self Organizing 
Information ``  ``. gr-qc/0009023 

30- M. Pavsic : Class. Quant. Gravity {\bf 7} (1990 ) L-187. 

M. Pavsic : Phys. Letts {\bf B 205}  ( 1998) 231. 

Phys. Letts {\bf B 221} ( 1999) 264.

M. Pavsic : Nuc. Phys. {\bf B 57 } ( Proc. Suppl) ( 1997) 265-268  

M. Pavsic : Riv. Il Nuovo Cimento { \bf 110 A } ( 4 ) ( 1997) 369. 

31- G. Segre : `` Quantum Algorithmic Randomness `` Ph. D Thesis, quant-ph/0009009 v3 .  

32-I. Antoniadis, P. Mazur , E. Mottola : `` Fractal Geometry of Quantum Spacetime at Large Scales `` 
hep-th/9808070             :

33- R. Oeckl : `` Braided Quantum Field Theory  `` hep-th/9906225 v2.  

34- C. Castro : `` $W$ Geometry from Fedosov Deformation Quantization `` Jour. Geom. and Physics {\bf 33 } ( 2000 ) 173.

35- C. Castro, J. Mahecha : `` Comments on the Riemann Conjecture and Index Theory on Cantorian-Fractal spacetime `` 
hep-th/0009014

36- P. Bouwknegt, K. Schouetens : Phys. Reports {\bf 223} (1993) 183-270 

37-  M . Vasiliev : `` Higher Spin gauge Theories, Star products and AdS `` hep-th/9910096.

E. Sezgin, P. Sundell : Higher Spin $N = 8$ Supergravity in $ AdS_4$  `` hep-th/9910096.

38. C. Castro : On the Four Dimensional Conformal Anomaly, Fractal Spacetime and the Fine Structure Constant `` 
physics/00010072 . 

C. Castro : Jour. Chaos, Solitons and Fractals {\bf 12} ( 2001) 101-104.

39. E. Gozzi : NPB {\bf 57 } ( Proc. Suppl ) (1997) 223. 

40- G. Kalberman : `` Communication through an extra dimension `` gr-qc/9910063.

\bye